%% file: main.tex
\documentclass[conference,letterpaper]{IEEEtran}
\usepackage[ruled,vlined,linesnumbered]{algorithm2e}
\usepackage[noend]{algpseudocode}

\input{preamble}
\SetKwProg{myfn}{function}{:}{end}
\SetKwFunction{MaxFlow}{MaxFlow}

\title{Info-Clustering: An Efficient Algorithm by\\ Network Information Flow}
\author{Chung Chan, Ali Al-Bashabsheh and Qiaoqiao Zhou
	\thanks{C.\ Chan, and Q.\ Zhou are with the Institute of Network Coding at the Chinese University of Hong Kong, the Shenzhen Key Laboratory of Network Coding Key Technology and Application, China, and the	Shenzhen Research Institute of the Chinese University of Hong Kong (email: cchan@inc.cuhk.edu.hk).
	}
	\thanks{A. Al-Bashabsheh was with the Institute of Network Coding at the	Chinese University of Hong Kong. He is now with the Big Data and Brain Computing (BDBC) center at Beihang University, Beijing, China (e-mail: entropyali@gmail.com).}
	\thanks{The work described in this paper was supported by a grant from University Grants Committee of the Hong Kong Special Administrative Region, China (Project No. AoE/E-02/08), and supported partially by a grant from Shenzhen Science and Technology Innovation Committee (JSGG20160301170514984), the Chinese University of Hong Kong (Shenzhen), China.}
	\thanks{The work of C.\ Chan was supported in part by The Vice-Chancellor's One-off Discretionary Fund of The Chinese University of Hong Kong (Project Nos. VCF2014030 and VCF2015007), and a grant from the University	Grants Committee of the Hong Kong Special Administrative Region, China (Project No. 14200714).}}
\begin{document}

\IEEEoverridecommandlockouts
\maketitle

\begin{abstract}
Motivated by the fact that entities in a social network or biological system often interact by exchanging information, we propose an efficient info-clustering algorithm that can group entities into communities using a parametric max-flow algorithm. This is a meaningful special case of the info-clustering paradigm where the dependency structure is graphical and can be learned readily from data. 
\end{abstract} 



\section{Introduction}

%
%
%

Info-clustering was proposed in~\cite{chan16cluster} as an application of network information theory
to the problem of clustering in machine learning. It regards each object as a piece of information,
namely a random variable, and groups random variables with sufficiently large amount of mutual
information together. Clustering is often an important first step in studying a large biological
system such as the human connectome and genome. It can also identify communities in a social network
so that resources can be allocated efficiently based on the communities discovered.
Since entities in a social or biological system often possess information and interact with each
other by the transmission of information, clustering them by their mutual information intuitively
gives meaningful results.

Using the multivariate mutual information (MMI) in \cite{chan15mi} as the similarity measure,
info-clustering defines a hierarchy of clusters of possibly different sizes at different levels of
mutual information. The clustering solution is intimately related to the principal sequence of partitions
(PSP)~\cite{narayanan90} of a submodular function, namely, that of the entropy function of the set of random
variables to be clustered. From this, it follows that the clustering solution is unique and solvable
using a polynomial number of oracle calls to evaluate the entropy function. However, in general,
learning the entropy function of a finite set $V$ of random variables from data takes exponential
time in the size of $V$~\cite{wu16}. The computation of the
PSP~\cite{narayanan90,nagano10}, \cite[Algorithm~3]{chan16cluster} without any approximation also
takes $\Omega(\abs {V}^2)$ calls to a submodular function minimization (SFM) algorithm, which in
turn makes $\Omega(\abs{V}^5)$ oracle calls to evaluate the submodular function. Hence, the
practicality of the general info-clustering algorithm is limited by both the sample complexity and
computational complexity. 

Fortunately, info-clustering reduces to faster algorithms under special statistical models that are
also easier (compared to a general model) to learn from data. For instance, under the Markov tree
model, info-clustering reduces to an edge-filtering procedure that runs in $O(|V|^2)$
time~\cite{chan15allerton}. Furthermore, this procedure coincides with an existing functional
genomic clustering method by mutual information relevance networks (MIRN)~\cite{butte00}. While
rediscovering a simple clustering algorithm under the Markov tree simplification, the info-clustering paradigm provides a theoretical
justification of the MIRN algorithm and helps discover how the algorithm may fail when the Markov tree
assumption does not hold~\cite[Example~6]{chan16cluster}.

In this work, we propose an efficient info-clustering algorithm under a different graphical model
called the pairwise independent network (PIN)~\cite{nitinawarat-ye10,nitinawarat10}. Using the idea
of the matroidal network link model in \cite{chan12ud}, the MMI has a concrete operational meaning
as the maximum network broadcast throughput~\cite{chan11isit}. The info-clustering solution
therefore identifies clusters with large intra-cluster communication rates, which naturally maps to
communities of closely related entities in a social network. Learning the PIN model simplifies to
learning the weights of $O(|V|^2)$ edges in a graph on $V$. In a social network, the weight
of each edge can simply be the amount/rate of communication between the edge's incident nodes.

As shown in \cite[Proposition~9]{chan16cluster}, the info-clustering solution for the PIN model can
be obtained from the PSP of the cut function of a weighted graph. It is well-known that faster SFM
algorithms are possible for the cut function using min-cut or max-flow algorithms~(e.g.,
see~\cite{fujishige1999minimizing,queyranne1998minimizing,jegelka2011fast}). An algorithm was given
in \cite{kolmogorov10} that computes the PSP efficiently by reducing the problem to a
parametric max-flow problem, where the capacities of the edges are certain monotonic functions of a
parameter. The reduction is carefully done such that the parametric max-flow algorithm in
\cite{gallo89} can compute the PSP in $O(\abs {V}^3\sqrt {\abs {E}})$ time, where $E$ is the
set of edges with non-zero weight. We will adapt this algorithm to compute the info-clustering
solution and modify it to improve the performance further.


\section{Preliminaries on info-clustering}

\subsection{Formulation}

\begin{figure*}
	\begin{center}
		\subcaptionbox{Weighted graph $G$ that represents of the PIN defined in~\eqref{eq:eg:src}.\label{fig:eg:src}}{
			\input{eg_src.pgf}
		}\hfill
		\subcaptionbox{Broadcast of $2$ message bits $\Rm_1$ and $\Rm_2$ with capacities defined in \eqref{eq:eg:c}.\label{fig:eg:I}}{
			\input{eg_I.pgf}
		}\hfill
		\subcaptionbox{The info-clustering solution in \eqref{eq:eg:clusters}.\label{fig:eg:clusters}}{
			\input{eg_clusters.pgf}
		}
	\end{center}
	\caption{Identifying the clusters of the PIN in~\eqref{eq:eg:src} by brute-force search over subsets satisfying the threshold constraint~\eqref{eq:cluster}.}
\label{fig:eg:PSP}
\end{figure*}

Let $V$ be the finite index set of the objects we want to cluster. Without loss of generality, we assume
\begin{align*}
  V &=[\abs {V}]:=\Set{1,\dots,\abs{V}}
\kern1em \text {and}\kern1em \abs{V}>1  \label{eq:V}
\end{align*}
The idea of info-clustering is to treat each object $i\in V$ as a random variable $\RZ_i$ (denoted in san serif font) taking values from a finite set $Z_i$ (denoted in the usual math font), and cluster the objects according to their mutual information. $P_{\RZ_V}$ denotes the distribution of the entire vector of random variables 
\begin{align*}
  \RZ_V &:=(\RZ_i\mid i\in V). \label{eq:RZV}
\end{align*}
We will illustrate the idea of info-clustering via a simple example shown in \figref{fig:eg:src}, where $V=[3]=\Set{1,2,3}$ and the corresponding random variables are defined as
\begin{subequations}
\label{eq:eg}
\begin{align}
	\label{eq:eg:src}
	\begin{split}
	\RZ_1&:=(\RX_{\rm{a}},\kern1.6em\RX_{\rm{c}})\\
	\RZ_2&:=(\RX_{\rm{a}},\RX_{\rm{b}}\kern1.7em)\\
	\RZ_3&:=(\kern1.7em\RX_{\rm{b}},\RX_{\rm{c}}),
	\end{split}
\end{align}
with $\RX_{\rm{a}},\RX_{\rm{b}}$ and $\RX_{\rm{c}}$ being independent uniformly random variables with entropies
\begin{align}
	\begin{split}
	H(\RX_{\rm{a}})&=H(\RX_{\rm{b}})=1\\ 
	H(\RX_{\rm{c}})&=5.
	\end{split}
\end{align}
This is a PIN~(Definition~\ref{def:PIN}) with correlation represented by the weighted triangle $G$ shown in \figref{fig:eg:src} characterized by the weight function $c$ where 
\begin{align}
	\label{eq:eg:c}
	\begin{split}
	c(\Set{1,2})&=c(\Set{2,3}):=H(\RX_{\rm{a}})=H(\RX_{\rm{b}})=1\\
	c(\Set{1,3})&:=H(\RX_{\rm{c}})=5.
	\end{split}
\end{align}
The vertex set is
$V:=[3]$ and the edge set is
\begin{align}
  \label{eq:eg:mcE}
  \mcE&:=\op{supp}(c)=\Set {\Set {1,2},\Set {2,3}, \Set {1,3}}.
\end{align}
\end{subequations}
Note that, for ease of comparison, this is the same graph used as an example in \cite{kolmogorov10} to illustrate the algorithm. A formal definition of the (hyper-)graphical source model is as follows:
\begin{Definition}[Definition~2.4 of \cite{chan10md}]
	\label{def:hyp}
	$\RZ_V$ is a \emph{hypergraphical source} w.r.t.\  a hypergraph $(V,E,`x)$ with edge functions $`x: E\to2^V`/\{\emptyset\}$ iff, for some independent (hyper)edge variables $\RX_e$ for $e\in E$ with $H(\RX_e)>0$,
	\begin{equation}
	\RZ_i:=(\RX_e\mid  e\in E, i\in`x(e)), \text{ for } i\in V. \label{eq:Xe}
	\end{equation}
	The \emph{weight function} $c:2^V`/\{\emptyset\}\to\mathbb{R}$ of a hypergraphical source is defined as
	\begin{subequations}
		\label{eq:c}
		\begin{align}
			c(B)&:=H(\RX_e\mid e\in E,`x(e)=B) \kern.5em \text{with support}\kern-.5em\\
			\kern-.5em \op{supp}(c)&:=\Set*{B\in 2^V`/\{\emptyset\} \mid c(B)>0}
		\end{align}
	\end{subequations}
\end{Definition}
The PIN model~\cite{nitinawarat10} is an example, where the corresponding hypergraph is a graph.
\begin{Definition}[\cite{nitinawarat10}]
	\label{def:PIN}
	$\RZ_V$ is a pairwise independent network (PIN) iff it is hypergraphical w.r.t.\  a graph $(V,E,`x)$ with edge function $`x: E\to V^2`/\{(i,i)\mid i\in V\}$ (i.e., no self loops).
\end{Definition}

The mutual information among multiple random variables is measured by the multivariate mutual information (MMI) defined in~\cite{chan15mi} as
\begin{subequations}
	\label{eq:mmi}
	\begin{align}
		I(\RZ_V) & := \min_{\mcP\in\Pi'(V)}I_{\mcP}(\RZ_V), \text{ with }\label{eq:I}\\
		I_{\mcP}(\RZ_V) & :=\frac{1}{|\mcP|-1}\biggl[\underbrace{\sum\nolimits_{C\in\mcP}H(\RZ_C)-H(\RZ_V)}_{=D(P_{\RZ_V}\|\prod_{C\in \mcP} P_{\RZ_C})}\biggr]\label{eq:IP}
	\end{align}
\end{subequations}
and $\Pi'(V)$ being the set of partitions of $V$ into at least 2 non-empty disjoint subsets of $V$. We may also write $I_{\mcP}(\RZ_V)$ more explicitly as 
\begin{align}
	I_{\mcP}(\RZ_V) = I(\RZ_{C_1}\wedge \dots \wedge \RZ_{C_k})\label{eq:IP:a}
\end{align}
for $\mcP=\Set{C_1,\dots,C_k}$. Note that Shannon's mutual information
$I(\RZ_1\wedge \RZ_2)$ is the special case when $\mcP$ is a
\emph{bipartition}. It is sometimes convenient to expand $I_{\mcP}(\RZ_V)$ using Shannon's mutual information~\cite[(5.18)]{chan15mi} as follows: 
\begin{align}
	I_{\mcP}(\RZ_V)&=I(\RZ_{C_1}\wedge \cdots \wedge C_k) \notag\\
	&= \frac1{k-1}\sum_{i=1}^{k-1} I(\RZ_{C_i} \wedge \RZ_{\bigcup_{j=i+1}^k C_j}). \label{eq:IPchain}
\end{align}
For the example with the random vector defined in \eqref{eq:eg:src},
\begin{subequations}
\label{eq:eg:I}
\begin{align}
\label{eq:eg:I2}
\begin{split}
	I(\RZ_{\Set{v,w}})&=I(\RZ_v\wedge \RZ_w)=c(\Set {v,w})\\
	&= \begin{cases}
		1, & \Set{v,w}\in \Set{\Set{1,2},\Set {2,3}}\\
		5, &  \Set{v,w}=\Set{1,3}
	\end{cases}
\end{split}
\end{align}
which reduces to Shanon's mutual information, and
\begin{align}
\label{eq:eg:I3}
\begin{split}
	I(\RZ_{[3]}) &=\min
	\big\{ I_{\Set{\Set{2,3},\Set{1}}}(\RZ_{[3]}),\\
	&\kern3.5em I_{\Set{\Set{1,3},\Set{2}}}(\RZ_{[3]}),\\
	&\kern3.5emI_{\Set{\Set{1,2},\Set{3}}}(\RZ_{[3]}),\\
	&\kern3.5emI_{\Set{\Set{1},\Set{2},\Set{3}}}(\RZ_{[3]})\big\}\\
	&=\min
	\big\{ I(\RZ_2,\RZ_3\wedge \RZ_1),\\
	&\kern3.5em I(\RZ_1,\RZ_3\wedge \RZ_2),\\
	&\kern3.5em I(\RZ_1,\RZ_2\wedge \RZ_3),\\
	&\kern3.5em \tfrac{I(\RZ_1,\RZ_3\wedge \RZ_2)+I(\RZ_1\wedge \RZ_3)}2\big\}\\
	&=\min \Set*{6,6,2,\tfrac{2+5}2} = 2,
\end{split}
\end{align}
\end{subequations}
where we have applied \eqref{eq:IPchain} to calculate $I_{\Set{\Set{1},\Set{2},\Set{3}}}(\RZ_{[3]})$ for the partition into singletons as the average of the value $I(\RZ_1,\RZ_3\wedge \RZ_2)$ of the cut that separates node $2$ from nodes $1$ and $3$, and the value $I(\RZ_1\wedge \RZ_3)$ of the cut that further separates node $1$ from node $3$.

Note that the sequence of two cuts effectively partitions the vertex set into singletons. From this expansion, it is clear that the partition into singletons cannot be optimal in this case, since the mutual information between nodes $1$ and $3$ is very large. Indeed, the optimal partition turns out to be a clustering of the random variable into correlated groups. In general, the set of optimal partitions to \eqref{eq:I}, denoted as $\Uppi^*(\RZ_V)$, form a semi-lattice w.r.t.\ the partial order that $\mcP\preceq \mcP'$ for the
partitions $\mcP$ and $\mcP'$ when
\begin{align}
	\forall C\in \mcP, \;\exists C'\in \mcP' : C\subseteq C'.\label{eq:<P}
\end{align}
$\mcP\prec \mcP'$ denotes the strict inequality. There is a unique finest/minimum partition $\mcP^*(\RZ_V)$, referred to as
the \emph{fundamental partition} for $\RZ_V$~\cite[Theorem~5.2]{chan15mi}. 
For a \emph{threshold} $`g\in `R$, the set of \emph{clusters} is defined as~\cite[Definition~1]{chan16cluster}
\begin{align}
	\pzC_{`g}(\RZ_V):= \op{maximal}\Set{B\subseteq V \mid \abs{B}>1, I(\RZ_B)>`g}\label{eq:cluster}
\end{align}
where $\op{maximal} \mathcal{F}$ is used to denote the inclusion-wise maximal elements of
$\mathcal{F}$, i.e.,
\begin{align}
	\op{maximal} \mcF := \left\{ B \in \mcF \mid \not\exists B'\supsetneq B, B'\in \mcF \right\}.\label{eq:maximal}
\end{align}

\begin{Proposition}[\mbox{\cite[Theorem~5]{chan16cluster}}]
\label{pro:firstcluster}
  $\pzC_{`g}(\RZ_V)=\mcP^*(\RZ_V)`/\Set{i\mid i\in V}$ with $`g=I(\RZ_V)$, i.e., the non-singleton subsets in the fundamental partition are the maximal subsets (clusters) with MMI larger than that of the entire set.
\end{Proposition}
For the example, applying the definition
~\eqref{eq:cluster} of clusters with the MMI calculated in \eqref{eq:eg:I}, the clustering solution is
\begin{align}
	\label{eq:eg:clusters}
	\pzC_{`g}(\RZ_{[3]})&=
	\begin{cases}
	\Set{[3]} & `g\in (-`8,2)\\ 
	\Set{1,3} & `g\in [2,5)\\ 
	`0 & `g\in [5,`8).
	\end{cases}
\end{align}
The info-clustering solution above consists of two clusters shown in \figref{fig:eg:clusters} for different intervals of the threshold $`g$. For $`g\geq 2$, the subset $\Set{1,3}$ is the only feasible solution that satisfies the threshold constraint in \eqref{eq:cluster}. For $`g\leq 2$, the entire set $[3]$ also satisfies the threshold constraint and is maximal. Recall from \eqref{eq:eg:I3} that there is a unique optimal partition for $I(\RZ_{\Set{1,2,3}})$, which is therefore the finest optimal partition
\begin{align}
  \mcP^*(\RZ_{[3]}) = \underbrace{\big\{\Set {1,3},}_{\pzC_{2}(\RZ_{[3]})} \Set {2}\big\}.
\end{align}
As expected from Proposition~\ref{pro:firstcluster}, the non-singleton element $\Set{1,3}$ is the only possible subset with MMI larger than $2$.

It turns out that the computation of the MMI, fundamental partition, and the entire info-clustering solution can be done in strongly polynomial time from the \emph{principal sequence of partition} (PSP) of the entropy function
\begin{align}
h:B\subseteq V\mapsto H(\RZ_B).\label{eq:h}
\end{align}
The PSP is a more general mathematical structure~\cite{narayanan90} in combinatorial optimization defined for a submodular function. More precisely, a reveal-valued set function $g:2^V\to `R$ is said to be \emph{submodular} iff
\begin{align}
  g(B_1)+g(B_2) \geq g(B_1\cap B_2)+g(B_1\cup B_2)\label{eq:submodular}
\end{align}
for all $B_1,B_2\subseteq V$. The entropy function, in particular, is a submodular function~\cite{fujishige78}.
The PSP of the submodular function $g$ is the characterization of the solutions to the following for all $`g\in `R$:
\begin{subequations}
\label{eq:DT}
\begin{align}
	\hat{g}(V):=\min_{\mcP\in \Pi(V)} g_{`g}[\mcP],\label{eq:DT:1}
\end{align}
referred to as the Dilworth truncation~\cite{schrijver02}, where $\Pi(V)$ is the partition of $V$ into one or more non-empty disjoint subsets,
\begin{align}
	g_{`g}[\mcP]&:=\sum_{C\in \mcP} g_{`g}(C)	\label{eq:g`gP}\\
        g_{`g}(C)&:=g(C)-`g 	\label{eq:g`g}
\end{align}
\end{subequations}
 (n.b., $\Pi'(V)$ in \eqref{eq:I} is $\Pi(V)$ but without the trivial partition $\Set{V}$, i.e., $\Pi'(V)=\Pi(V)`/\Set{\Set{V}}$.)
For every $`g$, submodularity of $g$ implies that there exists a unique finest/minimum (w.r.t.\ the partial order~\eqref{eq:<P}) optimal
partition to \eqref{eq:DT:1}, denoted as $\mcP^*(`g)$. It can be characterized as
\begin{subequations}
\label{eq:PSP}
\begin{align}
\label{eq:P*`g}
	\mcP^*(`g)=\mcP_{\ell} \kern1em \forall \gamma\in [`g_{\ell},`g_{\ell+1}), \ell \in \{0,...,N\}
\end{align}
for some integer $N>0$, a sequence of critical values of $`g$
\begin{align}
\label{eq:`gi}
	-`8 < \gamma_1<\dots<\gamma_N<`8
\end{align}
with $\gamma_0:=-`8$ and $\gamma_{N+1}:=`8$ for convenience, and a sequence of successively finer partitions
\begin{align}
\label{eq:Pi}
	\mcP_0={V}\succ \mcP_1\succ \dots \succ \mcP_N=\Set{\Set{i}|i\in V}.
\end{align}
\end{subequations}
The sequence of partitions (together with the corresponding critical values) is referred to as the PSP of $g$.
The PSP of the entropy function~\eqref{eq:h} characterizes the info-clustering solution as follows:

\begin{Proposition}[\mbox{\cite[Corollary~2]{chan16cluster}}]
	\label{pro:PSP}
	For a finite set $V$ with size $\abs{V}>1$ and a random vector $\RZ_V$,
	\begin{align}
		\begin{split}
	\pzC_{`g}(\RZ_V)
	&=`1[\min \Set{\mcP\in \Pi(V) \mid h_{`g}[\mcP]=\hat{h}_{`g}(V)}`2]\\ &\kern4em \big\backslash`1\{\Set{i}\mid i\in V`2\}, 
	   \end{split}
	\end{align}
namely, the non-singleton elements of the finest optimal partition to the Dilworth truncation~\eqref{eq:DT:1}.
\end{Proposition}


\section{Information flow interpretation}

For PIN, the MMI can be interpreted as the maximum broadcast throughput of a
network~\cite{chan11isit,chan12ud}, and hence info-clustering reduces to clustering by network information flow.
When applied to clustering social network, it can identify communities naturally based on the amount of information flow. 

More precisely, treating each edge as an undirected communication link with capacity $c$, at most a
total of $1$ bit can be communicated between node 1 and 2, and between node 2 and 3; and at most a total of $5$ bits
can be communicated between node 1 and 3. It can be seen that, for every pair of distinct
nodes $v,w \in V$, the broadcast throughput between $v$ and $w$ is given by the MMI in \eqref{eq:eg:I2}.
\figref{fig:eg:I} illustrates how $2$ bits of information can be broadcast in the entire network,
achieving the MMI of the entire set of random variables in \eqref{eq:eg:I3}. 
With the interpretation of the MMI as information flow, a cluster at threshold $`g$ is therefore a
maximal subnetwork with broadcast throughput larger than $`g$. For instance, the cluster $\Set
{1,3}\in \pzC_2(\RZ_{\Set{1,2,3}})$ is the only subset of nodes on which the induced subnetwork has a
throughput exceeding $2$.

\begin{figure*}
	\begin{center}
		\subcaptionbox{Weighted digraph $D$ by orienting $G$ as \eqref{eq:orient}.\label{fig:eg:orient}}{
			\input{eg_orient.pgf}
		}\hfill
		\subcaptionbox{The non-zero values of the incut function $g$ in \eqref{eq:eg:incut}.\label{fig:eg:cut}}{
			\input{eg_cut.pgf}
		}\hfill
		\subcaptionbox{The PSP~\eqref{eq:eg:P*} of $g$ and the Dilworth truncation from~\eqref{eq:eg:DT}.\label{fig:eg:P*}}{
		\input{eg_PSP.pgf}
		}
	\end{center}
	\caption{Computing the clusters of the PIN model~\eqref{eq:eg:src} as the non-singleton elements of the PSP~\eqref{eq:PSP}.}
	\label{fig:eg:PSP}
\end{figure*}
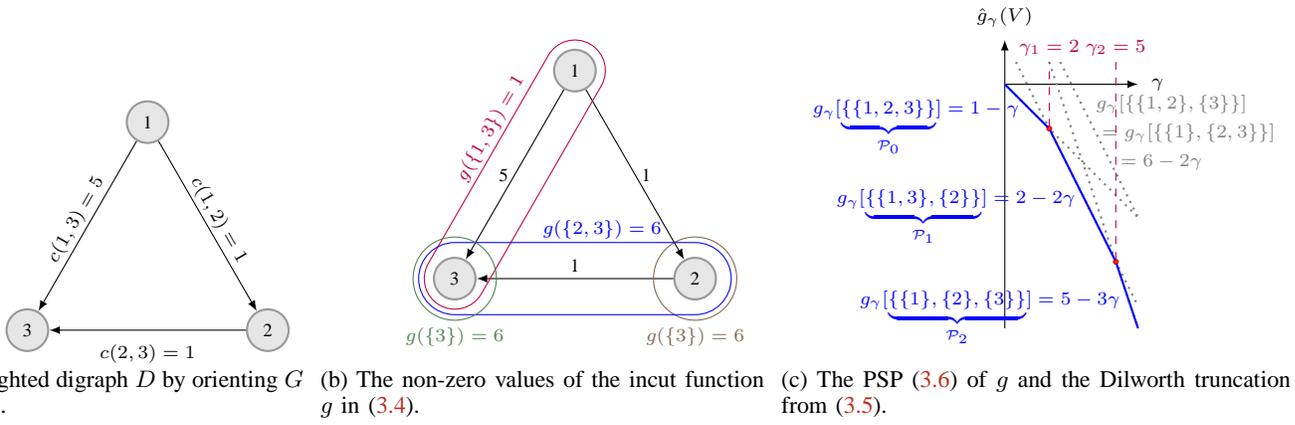

Specializing to the (hyper-)graphical model, it was shown in \cite[Proposition~8]{chan16cluster}
that the clustering solutions can be obtained directly as the non-singleton subsets from the PSP of
the incut function.
More precisely, from the weighted graph $G$ with vertex set $V$ and capacity function $c$, define
for every pair $(v,w)$ of vertices $v,w\in V$ 
\begin{align}
  \label{eq:orient}
  c(v,w)&:=
  \begin{cases}
    c(\Set{v,w}), & v<w\\
    0, & \text{otherwise.}
  \end{cases}
\end{align}
This defines the capacity function of a weighted digraph $D$ with vertex set $V$ and edge set $E$
which can be defined as the set of arcs $(v,w)$ with positive capacity $c(v,w)>0$. For example,
\figref{fig:eg:orient} is the weighted digraph obtained by orienting the weighted graph in
\figref{fig:eg:src} according to \eqref{eq:orient}, i.e., by directing an edge from the incident
node with a smaller label to the other incident node with a larger one.

For convenience, we also write for arbitrary subsets $B_1,B_2\subseteq V$
\begin{subequations}
\begin{alignat}{2}
	\label{eq:cut}
	c(B_1,B_2)&:=\sum_{v\in B_1} c(v,B_2)&\kern1em& \text{where}\\
        c(v,B_2)&:=\sum_{w\in B_2} c(v,w) && \text{for $v\in V$, and}\\
        c(B_1,w)&:=\sum_{w\in B_1} c(v,w) && \text{for $w\in V$.}
\end{alignat}
\end{subequations}
The incut function of the weighted digraph is defined as
\begin{align}
  g(B):=c(V`/B,B) \label{eq:g-PIN}.
\end{align}
The incut function for the digraph in \figref{fig:eg:orient} is shown in \figref{fig:eg:cut} and calculated below:
\begin{align}
\label{eq:eg:incut}
\begin{split}
	g(\Set{2})&=c(\Set{1,3},\Set{2})=c(1,2)+c(3,2)
	=1\\
	g(\Set{3})&=c(\Set{1,2},\Set{3})=c(1,3)+c(2,3)
	=6\\
	g(\Set{2,3})&=c(\Set{1},\Set{2,3})=c(1,2)+c(1,3)
	=6\\
	g(\Set{1,3})&=c(\Set{2},\Set{1,3})=c(2,1)+c(2,3)
	=1\\
	g(\Set {1})&=g(\Set {1,2})=g(\Set {1,2,3})=0.
\end{split}
\end{align}
To compute the PSP of $g$, we first evaluate \eqref{eq:g`gP} for different partitions as follows:
\begin{subequations}
	\label{eq:eg:DT}
\begin{align}
\begin{split}
g_{`g}[\Set{\Set{1,2,3}}]&=g_{`g}(\Set{1,2,3}) \\
&= g([3])-`g \\
&= -`g\\
g_{`g}[\Set{1,3},\Set{2}]&=g_{`g}(\Set{1,3})+g_{`g}(\Set{2}) \\
&= g(\Set{1,3})+g(\Set{2})-2`g\\
&=2-2`g\\
g_{`g}[\Set{1},\Set{2},\Set{3}]&=g_{`g}(\Set{1})+g_{`g}(\Set{2})+g_{`g}(\Set{3})\\
&= 7-3`g
\end{split}
\end{align} 
Similarly,
\begin{align}
\begin{split}
g_{`g}[\Set{1,2},\Set{3}]&=6-2`g >g_{`g}[\Set{1,3},\Set{2}]\\
g_{`g}[\Set{1},\Set{2,3}]&=6-2`g >g_{`g}[\Set{1,3},\Set{2}].
\end{split}
\end{align}
\end{subequations}
\figref{fig:eg:P*} plots the Dilworth truncation $\hat{g}_{`g}(V)$ in~\eqref{eq:DT} against $`g$ as
the minimum of $g_{`g}[\mcP]$ over all partitions $\mcP \in \Pi(V)$. It can be seen that for a given
$\mcP$, 
$g_{`g}[\mcP]$ is linear with integer slope $-\abs{\mcP}\in \Set{-|V|,\dots,1}$ and so $\hat{g}_{`g}(V)$
is piecewise linear consisting, in this example, of  $|V|=3$ line segments (highlighted in blue) and $|V|-1=2$
break points (highlighted in red). (In general, the number of line segments is at most $|V|$.) The
finest optimal partition for each value of $`g$ is 
\begin{align}
\label{eq:eg:P*}
\mcP^*(`g) &= 
\begin{cases}
	\underbrace{\Set{\Set{1,2,3}}}_{\mcP_0}, & `g \in (-`8,2]\\
	\underbrace{\Set{\Set{1,3},\Set{2}}}_{\mcP_1}, &   `g \in [\underbrace{2}_{`g_1},5)\\
	\underbrace{\Set{\Set{1},\Set{2},\Set{3}}}_{\mcP_2}, &   `g \in [\underbrace{5}_{`g_2},`8)\\
\end{cases}
\end{align}
with the PSP and the corresponding critical values annotated above and in the \figref{fig:eg:P*}.

\section{Clustering using parametric max-flow} 

By \cite[Proposition~9]{chan16cluster}, the PSP of the incut function $B\mapsto c(V`/B,B)$ of the
digraph $D$ coincides with the PSP of the cut function (divided by $2$) of the corresponding
undirected graph $G$, which was shown in \cite{kolmogorov10} to be computable by running a
\emph{parametric max-flow algorithm} $O(\abs {V})$ times. The parametric max-flow algorithm was
introduced by \cite{gallo89}, which runs in $O(\abs{V}^2 \sqrt{\abs{\mcE}})$ times using the
well-known \emph{push-relable}/\emph{preflow} algorithm~\cite{goldberg87-thesis,goldberg88-maxflow}
implemented with the highest-level selection rule~\cite{cherkassky97-push-relable}. Hence, the
info-clustering algorithm solution for the PIN model can be obtained in
$O(\abs{V}^3\sqrt{\abs{\mcE}})$ time. 

In this section, we will adapt and improve the algorithm in \cite{kolmogorov10} to compute the
desired PSP for the info-clustering solution. The algorithm will be illustrated using the same
example as in the last section, which is chosen to be the same example as in \cite{kolmogorov10} for
ease of comparison. 

We first give a procedure in Algorithm~\ref{alg:parametric-psp}
for computing the minimum minimizer $\mcP^*(`g)$ to \eqref{eq:DT} for all $`g\in `R$ and any
submodular function $g$, assuming a parametric submodular function minimizer. This procedure can be
specialized further to the PIN model where $g$ is chosen to be the incut function~\eqref{eq:g-PIN},
so that the parametric max-flow algorithm can be applied instead. 

\begin{algorithm}[h!]
	\caption{Computing the PSP as a parametric SFM.}
	\label{alg:parametric-psp}
	\DontPrintSemicolon
	\SetAlgoLined
	\KwIn{Submodular function $g:2^V\mapsto `R$ defined on a finite ground set $V$.}
	\KwOut{The function $\mcP^*(`g)$ of $`g\in `R$ defined in \eqref{eq:P*`g}.}
	$\mcP^*\leftarrow \Set{\Set{1}}$,
	$ B^*(`g)\leftarrow \Set{1}$, $x_{`g,1}\leftarrow g_{`g}(\Set{1})$,
	$\mu_i\leftarrow -\infty$ for all $i \in V$;\label{ln:PSP:init} \;
	\For{$j = 2$ \KwTo $\abs {V}$}{
		set $B^*(`g)$ as the (inclusion-wise) minimum minimizer to
		\begin{align}
			\label{eq:paraSFM}
			\min_{B\subseteq {[j]}:j\in B} g_{`g}(B) - x_{`g}(B\backslash\Set{j}),
		\end{align}
		where $x_{`g}(C):=\sum_{i\in C} x_{`g,i}$ for convenience;\;
		remove every $C$ that intersects 		$B^*(`g)$ from $\mcP^*(`g)$;\label{ln:P*:1}\;
		add $B^*(`g)$ to $\mcP^*(`g)$\label{ln:P*:2};\;
		\If{$j<|V|$}{
			\For{$i=1$ \KwTo $j$}{
				$\mu_i\leftarrow \max\Set*{\mu_i,\min\Set{`g\mid i\not\in B^*(`g)}}$ 
				\label{ln:`m}
				;\;
				$x_{`g,i}\leftarrow g_{\max\Set{`g,`m_i}}(\Set{i})$
				\label{ln:x}
				;\;
			}
		}
	}
\end{algorithm}

\begin{figure}[t!]
	\begin{center}
		
		\subcaptionbox{For the loop with $j=2$.\label{fig:eg:pm2}}{
			\kern1.2em\input{eg_pm2.pgf}
		}\\
		\subcaptionbox{For the loop with $j=3$.\label{fig:eg:pm3}}{
			\input{eg_pm3.pgf}
		}
	\end{center}
	\caption{Illustration of the computation of PSP in Algorithm~\ref{alg:parametric-psp}.}
	\label{fig:eg:psp}
\end{figure}
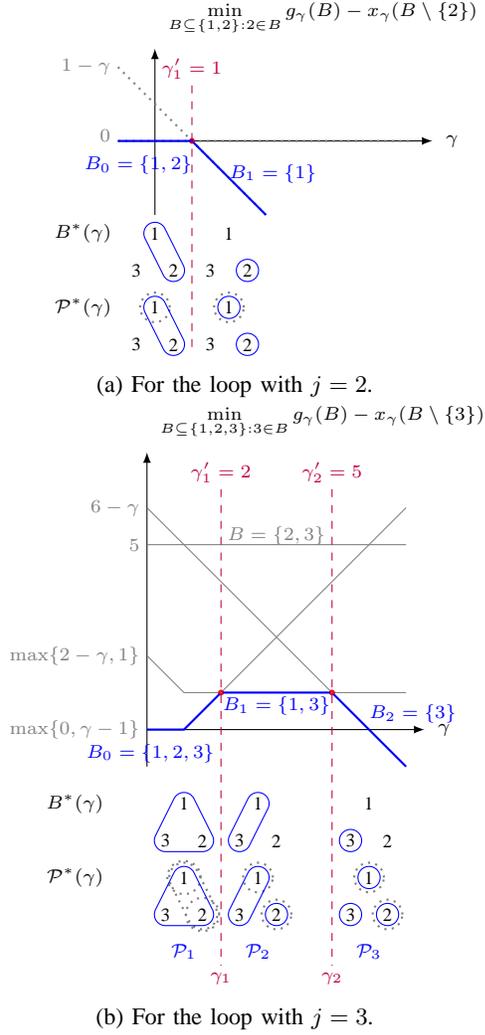

Consider the example with $g$ defined in \eqref{eq:eg:incut} and illustrated in \figref{fig:eg:cut}, and with $g_{`g}$ defined in \eqref{eq:g`g}.
When $j=2$, Line~\ref{ln:PSP:init} initializes $x_{`g,1}$ as
\begin{align}
	\label{eq:eg:x`g1}
	x_{`g,1} = g(\Set {1})-`g = -`g.
\end{align}
Then, \eqref{eq:paraSFM} becomes
\begin{align*}
	\min\Set {g_{`g}(\Set{1,2})-x_{`g,1},g_{`g}(\Set{2}) } 
	&= \min\Set {0,1-`g}\\
	&=\begin{cases}
		1-`g, & `g<1\\
		0 & `g\geq 1,
	\end{cases}
\end{align*}
which is a piecewise linear function plotted in 
\figref{fig:eg:pm2}. The minimum minimizer is therefore given by
\begin{align}
	\label{eq:eg:B^*:j=1}
	B^*(`g) &=
	\begin{cases}
		\underbrace{\Set {1,2}}_{B_0}, & `g<1\\
		\underbrace{\Set {2}}_{B_1}, & `g\geq \underbrace{1}_{`g'_1}.
	\end{cases}
\end{align} 
With $\mcP^*(`g)$ initialized in Line~\ref{ln:PSP:init} to $\Set {\Set {1}}$, Line~\ref{ln:P*:1}--\ref{ln:P*:2} update $\mcP^*(`g)$ to
\begin{align}
	\label{eq:eg:j=2:P*}
	\mcP^*(`g) = 
	\begin{cases}
		\big\{\underbrace{\Set {1,2}}_{B_0}\big\}, & `g<\underbrace{1}_{`g'_1}\\
		\big\{\Set {1},\underbrace{\Set {2}}_{B_1}\big\}, & `g\geq \underbrace{1}_{`g'_2}.
	\end{cases}
\end{align}
Next, with $`m_1$ and $`m_2$ initialized to $-`8$ in Line~\ref{ln:PSP:init}, the subsequent steps following Line~\ref{ln:P*:2} give
\begin{subequations}
	\label{eq:eg:x`m:j=3}
\begin{align}
	`m_1 &=\max\Set {-`8,`g'_1}=1\\
	x_{`g,1} &= g_{\max \Set {`g,1}}(\Set {1}) 
	= 
	\begin{cases}
		-1, & `g<1\\
		-`g, & `g\geq 1.
	\end{cases}\\
	`m_2 &=\max\Set {-`8,-`8}=-`8\\
	x_{`g,2} &= g_{\max \Set {`g,-`8}}(\Set {2}) = 1-`g.
\end{align}
\end{subequations}
Similarly, when $j=3$, the function of the minimum in \eqref{eq:paraSFM} is plotted in \figref{fig:eg:pm3}. It follows that the minimum minimizer is
\begin{align}
	\label{eq:eg:B^*:j=2}
	B^*(`g) &=
	\begin{cases}
		\underbrace{\Set {1,2,3}}_{B_0}, & `g \in (-`8,2]\\
		\underbrace{\Set {1,3}}_{B_1}, & `g\in [\underbrace{2}_{`g'_1},5)\\
		\underbrace{\Set {3}}_{B_2}, & `g\in [\underbrace{5}_{`g'_2},`8).
	\end{cases}
\end{align} 
With $\mcP^*(`g)$ given by \eqref{eq:eg:j=2:P*} and illustrated in \figref{fig:eg:pm2},
Lines~\ref{ln:P*:1}--\ref{ln:P*:2} update $\mcP^*(`g)$ to the desired solution~\eqref{eq:eg:P*}
characterized by the PSP.

The procedure is said to be parametric since the solution $\mcP^*$ is computed for all possible
 values of $`g\in `R$ rather than a particular value. To realize such a procedure, the characterization~\eqref{eq:P*`g} of $\mcP^*(`g)$ through the PSP in~\eqref{eq:Pi} and the corresponding critical values of $`g$ in \eqref{eq:`gi} is computed instead. 

The minimization in \eqref{eq:paraSFM} is a submodular function minimization (SFM) (over a lattice family). In
contrast with \cite{kolmogorov10}, we perform \eqref{eq:paraSFM} on a growing set $[j]=\Set{1,\dots,j}$ instead of the entire set $V$ in every loop. The idea follows from the algorithm for computing Dilworth truncation
such as the one given in \cite{schrijver02}. In contrast with \cite{schrijver02}, however, we follow
\cite{kolmogorov10} to
consider the update rule in Lines~\ref{ln:`m}--\ref{ln:x}, which guarantees $x_{`g,i}$ to be piecewise linear with at most one
break point, possibly at $`g=`m_i$ if $`m_i$ is finite. When $i=j$, the step in 
Line~\ref{ln:`m} gives $`m_j=-`8$ as $B^*(`g)$ always contain $j$ (see \eqref{eq:paraSFM}), and so
$x_{`g,j}=g_{`g}(\Set{j})$, which is linear without any breakpoint. As pointed out in \cite{kolmogorov10}, the update rule is particularly useful for the the parametric procedure since the complexity often grows with the number of break points. 

Specializing to PIN models where $g$ is the incut function defined in \eqref{eq:g-PIN}, the parametric SFM in~\eqref{eq:paraSFM} can be solved as a parametric min-cut problem as shown in Algorithm~\ref{alg:parametric-mc}.

\begin{algorithm}[h!]
	\caption{Computing the parametric SFM in~\eqref{eq:paraSFM} as a parametric min-cut.}
	\label{alg:parametric-mc}
	\DontPrintSemicolon
	\SetAlgoLined
	\KwIn{A weighted digraph $D$ on vertex set $V$ with capacity function $c:V^2\to `R$ satisfying \eqref{eq:orient}.}
	\KwOut{The minimum minimizer to \eqref{eq:paraSFM} with $g$ defined as the incut function in \eqref{eq:g-PIN}.}
	define a weighted digraph $D_j(`g)$ with vertex set $U\leftarrow \Set{s,1,\dots,j}$ (where $s$ is a new node outside $[j]$) and capacity function $c_{`g}:U^2\to `R$ initialized to $\M0$;\;
	\For{$v = 1$ \KwTo $j-1$}{
		$c_{`g}(s,v) \leftarrow \max\Set{0,-x_{`g,v}}$ \label{ln:c-gamma-s};\;
		$c_{`g}(v,j) \leftarrow \max\Set{0,x_{`g,v}}+c(v,j)$ \label{ln:c-gamma-j};\;
		$ c_{`g}(v,w) \leftarrow c(v,w)$ for all $w\in [j-1]`/[v]$ \label{ln:c-gamma-v};\;}
	compute the minimum minimizer $B^*(`g)$ to
	\begin{align}
		\min_{T\subseteq U\backslash\Set{s}: j\in T} c_{`g}(U\backslash T,T)
		\label{eq:paraMC}
	\end{align}
	which is the minimum $s$--$j$ cut value of $D_j(`g)$.
\end{algorithm}

\begin{figure*}
	\begin{center}
		\subcaptionbox{Weighted digraph $D_2(`g)$.\label{fig:eg:D2_digraph}}{
			\input{eg_D2.pgf}
		}\hfill
		\subcaptionbox{Construction of $D_2(`g)$ from $D$.\label{fig:eg:D2_contract}}{
			\input{eg_D2_contract.pgf}
		}\hfill
		\subcaptionbox{Minimum $s$--$2$ cut for $D_2(1)$.\label{fig:eg:D2_1}}{
			\input{eg_D2_1.pgf}
		}
	\end{center}
	\caption{Illustration of the parametric min-cut problem in Algorithm~\ref{alg:parametric-mc} for
the for-loop  with $j=2$.}
	\label{fig:eg:D2}
\end{figure*}
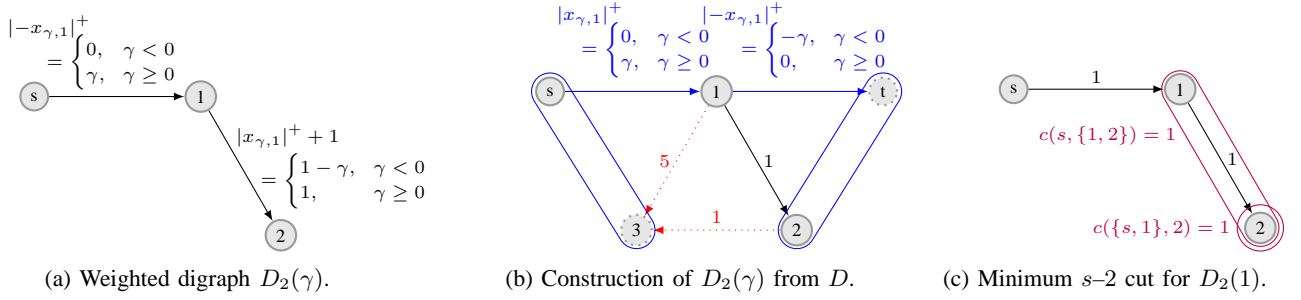

\begin{figure*}
	\begin{center}
		\subcaptionbox{Weighted digraph $D_3(`g)$.\label{fig:eg:D3_digraph}}{
			\input{eg_D3.pgf}
		}\hfill
		\subcaptionbox{Construction of $D_3(`g)$ from $D$.\label{fig:eg:D3_contract}}{
			\input{eg_D3_contract.pgf}
		}\hfill
		\subcaptionbox{$s$--$3$ max-flow and min-cut for $D_3(1)$.\label{fig:eg:D3_1}}{
			\input{eg_D3_1.pgf}
		}
	\end{center}
	\caption{Illustration of the parametric min-cut problem in Algorithm~\ref{alg:parametric-mc} for
the for-loop with $j=3$.}
	\label{fig:eg:D3}
\end{figure*}

For the current example, the digraph $D_j(`g)$ with $j=2$ is shown in \figref{fig:eg:D2}. The vertex set is $U=\Set {s,1,2}$.
The for-loop sets $v=1$ and initializes the capacity $c(s,1)=\abs {-x_{`g,1}}^+$ and $c(1,2)=\abs {x_{`g,1}}^++1$, where
\begin{align}
	\abs {x}^+ :=\max \Set {0,x}\kern1em \text {for $x\in `R$}.
\end{align}
Evaluating the capacities with the initial value of $x_{`g,1}$ in \eqref{eq:eg:x`g1} gives the non-decreasing and non-increasing piecewise linear functions $c(s,1)$ and $c(1,2)$ respectively shown in the figure. Similarly, the digraph $D_j(`g)$ with $j=3$ is shown in \figref{fig:eg:D3}, with the vertex set now being $U=\Set{s,1,2,3}$ instead, and $x_{`g,1}$ and $x_{`g,2}$ updated to the functions in \eqref{eq:eg:x`m:j=3}.

The weighted digraph $D_j(`g)$ can be viewed as a result of processing the weighted digraph $D$ as follow:
\begin{itemize}
	\item[1.] Augment the digraph $D$ with two new nodes $s$ and $t$ (both outside $V$) such that
	\begin{itemize}
		\item[a.] the capacity from $j$ to $t$ is set to infinity;
		\item[b.] for $v$ from $1$ to $j-1$, if $x_{`g,v}<0$, add an arc with capacity $-x_{`g,v}$
		from $s$ to $v$, else add an arc with capacity  $x_{`g,v}$ from $v$ to $t$.
	\end{itemize}
	\item[2.] Remove all outgoing arcs from node $j$ and contract node $t$ to node $j$.
	\item[3.] Remove all incoming arcs of nodes $j+1,...,|V|$ and contract the nodes to node $s$.
	\eqref{eq:paraMC}.
\end{itemize}
This procedure is illustrated for $D_2(`g)$ and $D_3(`g)$ in \figref{fig:eg:D2_contract} and \figref{fig:eg:D3_contract} respectively, where the red dotted arcs are removed, and the nodes circled together by blue lines are contracted.

%
Step~2 implies the formula in Line~\ref{ln:c-gamma-j} directly.
Step~3 gives
\begin{align*}
c_{`g}(s,v)=\max\Set{0,-x_{`g,v}}+\sum_{u\in V`/[j]} c(u,v)
\end{align*}
but this reduces to the formula in Line~\ref{ln:c-gamma-s} because the last summation is zero by the
assumption \eqref{eq:orient} that all the arcs point from a node with a smaller label to a node with
a larger label.
For the same reason, in Line~\ref{ln:c-gamma-v}, we do not need to set
$c_{`g}(v,w)=c(v,w)$ for $w\in [v-1]$ as they are zero by default. 
In contrast with \cite{kolmogorov10}, the additional node contraction and edge removal in Steps~2 and 3 above reduce the number of
vertices from $\abs {V}$ to $\abs {U}=j$
and therefore the complexity in solving \eqref{eq:paraMC}.


For each $j\in V$, \eqref{eq:paraMC} can be solved by the parametric max-flow algorithm in \cite{gallo89} in $O(j^3\sqrt{|E|})$ time by invoking $O(j)$ times the preflow algorithm \cite{goldberg87-thesis,goldberg88-maxflow} implemented with highest
level selection rule \cite{cherkassky97-push-relable}, which in turn
runs in $O(j^2\sqrt{|E|})$ times. The procedure is described in Algorithm~\ref{alg:parametric-mf}, which returns the characterization of the solution $B^*(`g)$ to \eqref{eq:paraMC} as 
\begin{align}
B^*(`g) = B_{\ell} \kern1em \text{ for } `g\in [`g'_{\ell},`g'_{\ell+1}), \ell \in \Set{0,...,N'}
\label{eq:B^*}
\end{align}
for some integer $N'>0$, where $`g'_{0}:=-`8, `g'_{N'+1}:=+`8$ and $B_{0}:=[j]$. Note also that $B^*(`g)=\Set{j}$ for sufficiently
large ${`g}$ and so $B_{N'+1}=\Set{j}$.

\begin{algorithm}[h!]
	\caption{Solving the parametric min-cut problem in \eqref{eq:paraMC} using the parametric max-flow algorithm~\cite{gallo89}.}
	\label{alg:parametric-mf}
	\DontPrintSemicolon
	\SetAlgoLined
	\KwIn{The weighted digraph $D_j(`g)$ on vertex set $U$ with capacity function $c_{`g}$ created in Algorithm~\ref{alg:parametric-mc}.}
	\KwOut{A list \DataSty{L} containing $(`g'_{\ell},B_{\ell})$ for $\ell \in [N']$ that characterizes \eqref{eq:paraMC} as in~\eqref{eq:B^*}.}
	create empty lists \DataSty{L} and \DataSty{PL};\;
	$
	`g^+\leftarrow \max_{v\in [j-1]}\Set{c([v-1],v)+c(v,[j]`/[v])}
	$;
	\label{ln:gamma-plus}\; 
	$
	`g^-\leftarrow \min\Set{\max\Set{`m_v,c(v,j)} \mid v\in [j-1]}
	$;\label{ln:gamma-minus}\; 
	\If{$`g^-=`g^+$}
	{
		\DataSty{L}$\leftarrow (`g^-,\Set{j})$ and return \DataSty{L};\label{ln:return1}\;
	}
	{
		set $f$ as the zero flow $\M0$ from $s$ to $j$;\;
		$[f^*,T^*]\leftarrow \MaxFlow(c_{`g^-},U,s,j,f)$;\label{ln:maxflow1}\;
		add $(`g^-,`g^+,f^*,\Set{s},\Set{j})$ to \DataSty{PL};\label{ln:addPL1}\;
	}
	\While{\DataSty{PL} is not empty}
	{
		withdraw any element $(`g^-,`g^+,f,S,T)$ from \DataSty{L};\label{ln:retrieve}\;
		compute $\bar{`g}\in [`g^-,`g^+]$ as the solution to 
		\begin{align}
			\label{eq:`gbar}
		c_{`g}(S,U`/S) = c_{`g}(U`/T,T);
		\end{align}
	\vspace{-1em}
		\;
		define a weighted digraph $\bar{D}_j$ with vertex set $\bar{U}\leftarrow ([j]`/(S\cup T)) \cup
		\Set{s,j}$ and capacity function $\bar{c}:\bar{U}^2\to `R$ initialized to $\M0$;\;
		\For{$v$ in $\bar{U}`/\Set{s,j}$}{
			$\bar{c}(s,v) \leftarrow \sum_{u\in S} c_{\bar{`g}}(u,v)$ \label{ln:barc-s};\;
			$\bar{c}(v,j) \leftarrow  \sum_{w\in T} c_{\bar{`g}}(v,w)$ \label{ln:barc-j};\;
			$\bar{c}(v,w) \leftarrow c(v,w)$ for all $w\in \bar{U}`/\Set{s,j,v}$ \label{ln:barc-v};\;
		}		
		\For{$v$ in $\bar{U}`/\Set{j}$}
		{
			$\bar{f}(v,j) \leftarrow  \sum_{w\in T} \min\Set{f(v,w),\bar{c}(v,j)}$ and $\bar{f}(j,v)\leftarrow -\bar{f}(v,j)$ to ensure anti-symmetry;\label{ln:barf-j}\;
			$\bar{f}(v,w) \leftarrow f(v,w)$ for all $w\in \bar{U}`/\Set{j,v}$ \label{ln:barf-v};\;
		}
		$[f^*,T^*]\leftarrow\MaxFlow(\bar{c},\bar{U},s,j,f)$;\label{ln:maxflow2}\;
		\If{$T^*=\Set{j}$}{
			add $(`g^-,T)$ to \DataSty{L};\label{ln:addL}\;
		}{
			add $(`g^-,\bar{`g},f,S,T\cup T^*)$ and $(\bar{`g},`g^+,f^*,S\cup (U`/T^*), T)$ to \DataSty{PL};\label{ln:addPL}\;}
	}
\end{algorithm}

To solve \eqref{eq:paraMC} for any fixed $j$, we assume the following subroutine
\begin{align}
	[f^*,T^*]=\MaxFlow(\bar{c},\bar{U},\bar{s},\bar{t},\bar{f})\label{eq:maxflow}
\end{align}
which takes as arguments the capacity
function $\bar{c}$ (fixed and not parametric), the vertex set $\bar{U}$ on which $\bar{c}$ is defined,
the source node $\bar{s}\in \bar{U}$, the sink node $\bar{t}\in \bar{U}$ and a valid preflow $\bar{f}$
associated with the weighted digraph $D_{j}(`g)$ defined by the previous arguments.
It returns the maximum $\bar{s}$--$\bar{t}$ flow $f^*$ and the inclusion-wise minimum set $T^*$ that solves
\begin{align}
	\label{eq:triple-star}
	\min_{T\subseteq \bar{U}`/\Set{\bar{s}}: \bar{t}\in T} c(\bar{U}`/T,T),
\end{align}
and is referred to as the minimum $\bar{s}$--$\bar{t}$ cut.

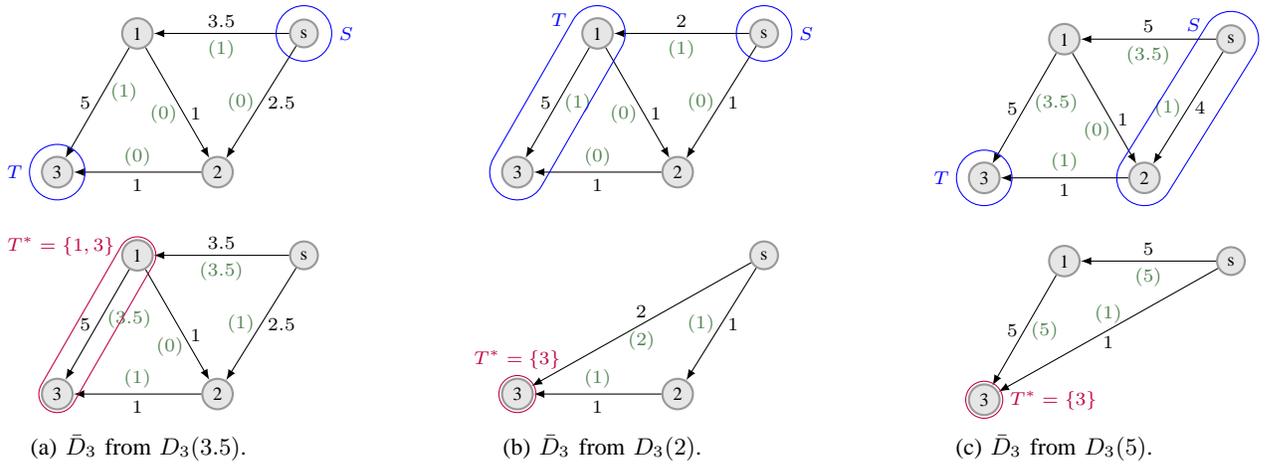
\begin{figure*}
	\begin{center}
		\subcaptionbox{$\bar{D}_3$ from $D_3(3.5)$.\label{fig:eg:D3_35}}{
			\input{eg_D3_35.pgf}
		}\hfill
		\subcaptionbox{$\bar{D}_3$ from $D_3(2)$.\label{fig:eg:D3_2}}{
			\input{eg_D3_2.pgf}
		}\hfill
		\subcaptionbox{$\bar{D}_3$ from $D_3(5)$.\label{fig:eg:D3_5}}{
			\input{eg_D3_5.pgf}
		}
	\end{center}
	\caption{Illustration of the parametric max-flow algorithm in Algorithm~\ref{alg:parametric-mf}.}
	\label{fig:eg:barD3}
\end{figure*}

Roughly speaking, $`g^+$ in Line~\ref{ln:gamma-plus} is the value of $`g$ at which
\eqref{eq:paraMC} (i.e., $B^{*}(`g)$) is constant for $`g\geq `g^+$.
Similarly, $`g^-$ in Line~\ref{ln:gamma-minus} is the value of $`g$ at which \eqref{eq:paraMC} is constant for $`g\leq `g^-$.
When $`g^-=`g^+$, there is only one critical value $`g'_1$ of $`g$ where $B^*(`g)$ changes from $B_0=[j]$ to $B_1=\Set {j}$. 

To illustrate the above, consider $j=2$, i.e., with $D_{2}(`g)$ shown in \figref{fig:eg:D2_digraph} as the input to
Algorithm~\ref{alg:parametric-mf}.
Then, Lines~\ref{ln:gamma-plus}--\ref{ln:gamma-minus} give
\begin{align*}
	`g^+&=c(1,2)=1\\
	`g^-&=\max\Set {`m_1,c(1,2)}=1,
\end{align*} 
where the last equality is because $`m_1$ is initialized to be $-`8$ by Line~\ref{ln:PSP:init} of
Algorithm~\ref{alg:parametric-psp}.
Since $`g^-=`g^+$ in this case, the algorithm returns at Line~\ref{ln:return1} the list
\begin{align*}
	\DataSty{L}=[(\underbrace{1}_{`g'_1},\underbrace{\Set {1}}_{B_1})].
\end{align*} 
This gives the desired $B^*(`g)$ in \eqref{eq:eg:B^*:j=1}. \figref{fig:eg:D2_1} shows the digraph
$D_2(`g)$ at $`g=1$. It can be seen that both $\Set {1,2}$ and $\Set {2}$ are solutions to the
minimization in \eqref{eq:paraMC}.

If $`g^-\neq `g^+$ (or more specifically $`g^-< `g^+$), then the interval $(`g^-,`g^+)$ must contain
other critical values of $`g$ where $B^*(`g)$ changes. The critical values are then computed
iteratively by the preflow algorithm $\MaxFlow$~\eqref{eq:maxflow} (Lines~\ref{ln:maxflow1} and
\ref{ln:maxflow2}) applied on the digraph $\bar{D}_j$ with capacities derived from those of
$D_j(`g)$ (Lines~\ref{ln:barc-s}--\ref{ln:barc-v}), and with $`g$ evaluated at some value $\bar
{`g}\in (`g^-,`g^+)$ satisfying \eqref{eq:`gbar}. This either resolves $B^*(`g)$ for the entire
interval (in which case the solution is updated in Line~\ref{ln:addL}) or reduces the problem to two
smaller subproblems for later processing (i.e., with the original interval $(`g^-,`g^+)$ replaced by
the two smaller intervals $(`g^-,\bar{`g})$ and $(\bar{`g},`g^+)$ in Line~\ref{ln:addPL}). 

To illustrate the procedure above, consider $j=3$, i.e., with $D_3(`g)$ shown in
\figref{fig:eg:D3_digraph}
as the input to Algorithm~\ref{alg:parametric-mf}. Then, Lines~\ref{ln:gamma-plus}--\ref{ln:gamma-minus} give
\begin{align*}
	`g^+ &= \max \Set { c(1,2)+c(1,3), c(1,2)+c(2,3) }\\
	&= \max \Set {1+5,1+1} = 6\\
	`g^-&= \min \Set {\max \Set {`m_1, c(1,3)}, \max \Set {`m_2, c(2,3)} }\\
	&= \min \Set {\max \Set {1, 5}, \max \Set {-`8,1} } = 1
\end{align*}
where we used the values $`m_1=1$ and $`m_2=-`8$ by \eqref{eq:eg:x`m:j=3}. Since $`g^-<`g^+$ in this
case, Line~\ref{ln:return1} is skipped. Line~\ref{ln:maxflow1} invokes the preflow algorithm for the
graph $D_3(`g^-)=D_3(1)$ shown in \figref{fig:eg:D3_1}. The min-cut is $T^*=[3]=U`/\Set {s}$ (where
$U=\Set {s,1,2,3}$ is the vertex set of $D_3$) by the construction of $`g^-$. The max-flow is
\begin{align}
	\begin{split}
		&f^*(s,1)=f^*(1,3)=1\\
		&f^*(1,s)=f^*(3,1)=-1
	\end{split}
\end{align}
and $0$ otherwise. Note that the second line of equations ensures the anti-symmetry property of a
flow function, i.e., 
\begin{align}
	f^*(w,v)=-f^*(v,w) \label{eq:anti-symmetry}
\end{align}
for all pairs of distinct nodes $v$ and $w$. The flow along each arc is indicated  in
\figref{fig:eg:D3_1} by the parentheses next to the corresponding capacity of the arc.

The tuple $(`g^-,`g^+,f^*,\Set{s},\Set {j})$ is then added to $\DataSty{PL}$ in Line~\ref{ln:addPL1}
and then retrieved (and deleted from \DataSty{PL}) subsequently inside the while-loop (Line~\ref{ln:retrieve}). 
With $`g^-=1$,
$`g^+=6$, $S=\Set {s}$ and $T=\Set {j}$ in Line~\ref{ln:retrieve},
the l.h.s.\ of \eqref{eq:`gbar} is given as (see \figref{fig:eg:D3_digraph})
\begin{align*}
	c_{`g}(S,U`/S) &= c_{`g}(s, \Set {1,2,3})\\ &= 
	\begin{cases}
		1, &`g<1\\
		`g, &`g\geq 1
	\end{cases}
	+
	\begin{cases}
		0, & `g<1\\
		`g-1, & `g\geq 1
	\end{cases}\\
&= 	\begin{cases}
	1, & `g<1\\
	2`g-1, & `g\geq 1
\end{cases}
\end{align*}
and r.h.s.\ of \eqref{eq:`gbar} is given as
\begin{align*}
	c_{`g}(U`/T,T) &= c_{`g}(\Set {s,1,2},3)\\ &= 
	5
	+
	\begin{cases}
		2-`g, & `g<1\\
		1, & `g\geq 1
	\end{cases}\\
	&= 	\begin{cases}
		7-`g, & `g<1\\
		6, & `g\geq 1.
	\end{cases}
\end{align*}
$\bar {`g}$ is computed as the solution to \eqref{eq:`gbar}, namely $\bar{`g}=3.5$. In general, such a value must exist and is unique because $U`/S$ and $T$ are optimal solutions to \eqref{eq:paraMC}
at $`g^-$ and $`g^+$ respectively. The computation is in $O(j)$ time since both sides of the
equations are piecewise linear with at most $O(j)$ break points.

The new weighted digraph $\bar{D}_j$ with the capacity function $\bar{c}$ assigned in the first
for-loop (Lines~\ref{ln:barc-s}--\ref{ln:barc-v}) can be obtained from $D_j(`g)$ by 
\begin{enumerate}
	\item setting $`g=\bar{`g}$, contracting $S$ to the source node $s$,
	\item contracting $T$ to the sink node $j$, and then
	\item removing the incoming
	arcs to $s$ and outgoing arcs from~$j$.
\end{enumerate}
 The second for-loop turns $f$ to a valid preflow $\bar{f}$ of $\bar{D}_j$. 

Recall that for the current example, $\bar {`g}=3.5$ in the last execution of the algorithm.
\figref{fig:eg:D3_35} shows two digraphs,
where the top one is the digraph $D_3(`g)$ at $`g=\bar{`g}=3.5$ and the bottom one is the new weighted digraph $\bar {D}_3$.
The sets $S$, $T$ and the flow $f$ indicated on the top
digraph $D_3(3.5)$ satisfy \eqref{eq:`gbar}, while $\bar {D}_3$ is annotated with the max-flow $f^*$
and min-cut $T^*$ computed by Line~\ref{ln:maxflow2}.  Note that, since $T^*=\Set{1,3}\neq \Set
{j}$, Line~\ref{ln:addL} will be skipped. Instead, Line~\ref{ln:addPL} adds the following two tuples
to the list $\DataSty{PL}$, which becomes
\begin{align}
	\DataSty{PL}=[(1,3.5,f^*,\Set {s}, \Set {1,3}),(3.5,6,f,\Set {s,2}, \Set {3})]. \label{eq:eg:PL}
\end{align}
Repeating the while-loop with the first element retrieved from \DataSty{PL}, it can be shown that \eqref{eq:`gbar} is solved by the value $\bar {`g}=2$. Similar to \figref{fig:eg:D3_35}, \figref{fig:eg:D3_2} shows the digraph $D_3(2)$ at the top and $\bar{D}_3$ at the bottom. It can be verified that $S$ and $T$ satisfies \eqref{eq:`gbar} for the top graph and $T^*$ is the min-cut in the bottom graph. Since $T^*=\Set {3}$ in this case, a new element $(2,\Set {1,3})$ is added to $\DataSty{L}$ in Line~\ref{ln:addL}. 

Finally, repeating the while-loop again with the last element retrived from \DataSty{PL}~\eqref{eq:eg:PL}, it can be shown that $\bar{`g}=5$. \figref{fig:eg:D3_5} again gives $D_3(5)$ and the min-cut $T^*=\Set {3}$, in which case a new element $(5,\Set {3})$ is added to \DataSty{L} again in Line~\ref{ln:addL}. Since $\DataSty{PL}$ is not empty, the algorithm terminates with
\begin{align*}
	\DataSty{L}=[(\underbrace{2}_{`g'_1},\underbrace{\Set {1,3}}_{B_1}),(\underbrace{5}_{`g'_2},\underbrace{\Set {3}}_{B_2})].
\end{align*} 
This gives the desired $B^*(`g)$ in \eqref{eq:eg:B^*:j=2} that yields the desired PSP in \eqref{eq:eg:P*}, and therefore the info-clustering solution in~\eqref{eq:eg:clusters} for the PIN model~\eqref{eq:eg:src}.

\section{Conclusion}
\label{sec:conclusion}

We have adapted the parametric max-flow algorithm of computing the PSP to an info-clustering algorithm that clusters a graphical network based on the information flow over its edges. The overall running time is $O(\abs {V}^3\sqrt {\abs {E}})$, where $\abs {V}$ is the size of the network and $\abs {E}$ is the number of edges or communication link. The algorithm simplifies the general info-clustering algorithm by a few orders of magnitude, and is applicable to systems, such as the social networks, where similarity can be measured by mutual information. 

To implement the algorithm in a large-scale social network, the preflow algorithm may be made distributive and adaptive: 
Servers may be deployed in different parts of the network to measure and store the information exchange rates of different pair of nodes. The push and relabel operations in the preflow algorithm can be done locally by the servers first and then communicated to other servers when necessary. 
The preflow of the network may be stored in conjunction with the clustering solution, so that the clusters can be updated incrementally over time based on the changes of information exchange rates. The allocation of the servers and other resources may also be adapted to the clustering solution. For instance, as intra-cluster communication is more frequent than inter-cluster communication, the nodes in a cluster with larger mutual information may be assigned to the same server so that changes in the network can be updated more frequently without much communication overhead among the servers.

\bibliographystyle{IEEEtran}
\bibliography{IEEEabrv,ref}
\end{document}

%% file: preamble.tex
\usepackage[numbers,sort&compress]{natbib}

\makeatletter

\usepackage{etex}

\usepackage{zref-savepos}

\newcounter{mnote}

\def\xmarginnote{%
  \xymarginnote{\hskip -\marginparsep \hskip -\marginparwidth}}

\def\ymarginnote{%
  \xymarginnote{\hskip\columnwidth \hskip\marginparsep}}

\long\def\xymarginnote#1#2{%
\vadjust{#1%
\smash{\hbox{{%
        \hsize\marginparwidth
        \@parboxrestore
        \@marginparreset
\footnotesize #2}}}}}

\def\mnoteson{%
\gdef\mnote##1{\refstepcounter{mnote}\label{##1}%
  \zsavepos{##1}%
  \ifnum20432158>\number\zposx{##1}%
  \xmarginnote{{\color{blue}\bf $\langle$\arabic{mnote}$\rangle$}}%
  \else
  \ymarginnote{{\color{blue}\bf $\langle$\arabic{mnote}$\rangle$}}%
  \fi%
}
  }
\gdef\mnotesoff{\gdef\mnote##1{}}
\mnoteson
\mnotesoff

\newcommand{\figref}[1]{Fig.~\ref{#1}}

\usepackage{framed}
\usepackage{subcaption}
\usepackage{comment}
\usepackage[svgnames,dvipsnames]{xcolor}
\usepackage[all]{xy}
\usepackage{tikz}
\usetikzlibrary{positioning,matrix,through,calc,arrows,fit,shapes,decorations.pathreplacing,decorations.markings,decorations.text}

\tikzstyle{block} = [draw,fill=blue!20,minimum size=2em]

\tikzset{%
	prefix node name/.code={%
		\tikzset{%
			name/.code={\edef\tikz@fig@name{#1 ##1}}
		}%
	}%
}

\@ifpackagelater{tikz}{2013/12/01}{
	\newcommand{\convexpath}[2]{
		[create hullcoords/.code={
			\global\edef\namelist{#1}
			\foreach [count=\counter] \nodename in \namelist {
				\global\edef\numberofnodes{\counter}
				\coordinate (hullcoord\counter) at (\nodename);
			}
			\coordinate (hullcoord0) at (hullcoord\numberofnodes);
			\pgfmathtruncatemacro\lastnumber{\numberofnodes+1}
			\coordinate (hullcoord\lastnumber) at (hullcoord1);
		}, create hullcoords ]
		($(hullcoord1)!#2!-90:(hullcoord0)$)
		\foreach [evaluate=\currentnode as \previousnode using \currentnode-1,
		evaluate=\currentnode as \nextnode using \currentnode+1] \currentnode in {1,...,\numberofnodes} {
			let \p1 = ($(hullcoord\currentnode) - (hullcoord\previousnode)$),
			\n1 = {atan2(\y1,\x1) + 90},
			\p2 = ($(hullcoord\nextnode) - (hullcoord\currentnode)$),
			\n2 = {atan2(\y2,\x2) + 90},
			\n{delta} = {Mod(\n2-\n1,360) - 360}
			in 
			{arc [start angle=\n1, delta angle=\n{delta}, radius=#2]}
			-- ($(hullcoord\nextnode)!#2!-90:(hullcoord\currentnode)$) 
		}
	}
}{
	\newcommand{\convexpath}[2]{
		[create hullcoords/.code={
			\global\edef\namelist{#1}
			\foreach [count=\counter] \nodename in \namelist {
				\global\edef\numberofnodes{\counter}
				\coordinate (hullcoord\counter) at (\nodename);
			}
			\coordinate (hullcoord0) at (hullcoord\numberofnodes);
			\pgfmathtruncatemacro\lastnumber{\numberofnodes+1}
			\coordinate (hullcoord\lastnumber) at (hullcoord1);
		}, create hullcoords ]
		($(hullcoord1)!#2!-90:(hullcoord0)$)
		\foreach [evaluate=\currentnode as \previousnode using \currentnode-1,
		evaluate=\currentnode as \nextnode using \currentnode+1] \currentnode in {1,...,\numberofnodes} {
			let \p1 = ($(hullcoord\currentnode) - (hullcoord\previousnode)$),
			\n1 = {atan2(\x1,\y1) + 90},
			\p2 = ($(hullcoord\nextnode) - (hullcoord\currentnode)$),
			\n2 = {atan2(\x2,\y2) + 90},
			\n{delta} = {Mod(\n2-\n1,360) - 360}
			in 
			{arc [start angle=\n1, delta angle=\n{delta}, radius=#2]}
			-- ($(hullcoord\nextnode)!#2!-90:(hullcoord\currentnode)$) 
		}
	}
}

\usepackage{qsymbols,amssymb,mathrsfs}
\usepackage{amsmath}
\usepackage[standard,thmmarks]{ntheorem}
\theoremstyle{plain}
\theoremsymbol{\ensuremath{_\vartriangleleft}}
\theorembodyfont{\itshape}
\theoremheaderfont{\normalfont\bfseries}
\theoremseparator{}

\theoremstyle{nonumberplain}
\theoremheaderfont{\scshape}
\theorembodyfont{\normalfont}
\theoremsymbol{\ensuremath{_\blacktriangleleft}}

\theoremnumbering{arabic}
\theoremstyle{plain}
\usepackage{latexsym}
\theoremsymbol{\ensuremath{_\Box}}
\theorembodyfont{\itshape}
\theoremheaderfont{\normalfont\bfseries}
\theoremseparator{}

\theorembodyfont{\upshape}
\theoremprework{\bigskip\hrule}
\theorempostwork{\hrule\bigskip}

\usepackage[overload]{empheq} 

\let\iftwocolumn\if@twocolumn
\g@addto@macro\@twocolumntrue{\let\iftwocolumn\if@twocolumn}
\g@addto@macro\@twocolumnfalse{\let\iftwocolumn\if@twocolumn}

\mathtoolsset{showonlyrefs=false,showmanualtags}
\let\underbrace\LaTeXunderbrace 

\renewcommand{\eqref}[1]{\textup{(\refeq{#1})}} 
\newtagform{brackets}[]{(}{)}   
\usetagform{brackets}

\usepackage[Smaller]{cancel}

\usepackage{graphicx,psfrag}
\graphicspath{{figure/}{image/}} 

\usepackage{diagbox} 

\usepackage{algorithm2e}
\usepackage{listings} 
\lstdefinelanguage{Maple}{
  morekeywords={proc,module,end, for,from,to,by,while,in,do,od
    ,if,elif,else,then,fi ,use,try,catch,finally}, sensitive,
  morecomment=[l]\#,
  morestring=[b]",morestring=[b]`}[keywords,comments,strings]
\DeclareMathAlphabet{\mathpzc}{OT1}{pzc}{m}{it}
\usepackage{upgreek} 
\usepackage{dsfont}  

\def\multi@nostar#1#2{%
  \expandafter\def\csname multi#1\endcsname##1{%
    \if ##1.\let\next=\relax \else
    \def\next{\csname multi#1\endcsname}     
    \expandafter\newcommand\csname #1##1\endcsname{#2}
    \fi\next}}

\def\multi@star#1#2{%
  \expandafter\def\csname #1\endcsname##1{#2}
  \multi@nostar{#1}{#2}
}

\newcommand{\multi}{%
  \@ifstar \multi@star \multi@nostar}

\multi*{rm}{\mathrm{#1}}
\multi*{mc}{\mathcal{#1}}
\multi*{op}{\mathop {\operator@font #1}}
\multi*{ds}{\mathds{#1}}
\multi*{set}{\mathcal{#1}}
\multi*{rsfs}{\mathscr{#1}}
\multi*{pz}{\mathpzc{#1}}
\multi*{M}{\boldsymbol{#1}}
\multi*{R}{\mathsf{#1}}
\multi*{RM}{\M{\R{#1}}}
\multi*{bb}{\mathbb{#1}}
\multi*{td}{\tilde{#1}}
\multi*{tR}{\tilde{\mathsf{#1}}}
\multi*{trM}{\tilde{\M{\R{#1}}}}
\multi*{tset}{\tilde{\mathcal{#1}}}
\multi*{tM}{\tilde{\M{#1}}}
\multi*{baM}{\bar{\M{#1}}}
\multi*{ol}{\overline{#1}}

\multirm  ABCDEFGHIJKLMNOPQRSTUVWXYZabcdefghijklmnopqrstuvwxyz.
\multiol  ABCDEFGHIJKLMNOPQRSTUVWXYZabcdefghijklmnopqrstuvwxyz.
\multitR   ABCDEFGHIJKLMNOPQRSTUVWXYZabcdefghijklmnopqrstuvwxyz.
\multitd   ABCDEFGHIJKLMNOPQRSTUVWXYZabcdefghijklmnopqrstuvwxyz.
\multitset ABCDEFGHIJKLMNOPQRSTUVWXYZabcdefghijklmnopqrstuvwxyz.
\multitM   ABCDEFGHIJKLMNOPQRSTUVWXYZabcdefghijklmnopqrstuvwxyz.
\multibaM   ABCDEFGHIJKLMNOPQRSTUVWXYZabcdefghijklmnopqrstuvwxyz.
\multitrM   ABCDEFGHIJKLMNOPQRSTUVWXYZabcdefghijklmnopqrstuvwxyz.
\multimc   ABCDEFGHIJKLMNOPQRSTUVWXYZabcdefghijklmnopqrstuvwxyz.
\multiop   ABCDEFGHIJKLMNOPQRSTUVWXYZabcdefghijklmnopqrstuvwxyz.
\multids   ABCDEFGHIJKLMNOPQRSTUVWXYZabcdefghijklmnopqrstuvwxyz.
\multiset  ABCDEFGHIJKLMNOPQRSTUVWXYZabcdefghijklmnopqrstuvwxyz.
\multirsfs ABCDEFGHIJKLMNOPQRSTUVWXYZabcdefghijklmnopqrstuvwxyz.
\multipz   ABCDEFGHIJKLMNOPQRSTUVWXYZabcdefghijklmnopqrstuvwxyz.
\multiM    ABCDEFGHIJKLMNOPQRSTUVWXYZabcdefghijklmnopqrstuvwxyz.
\multiR    ABCDEFGHIJKL NO QR TUVWXYZabcd fghijklmnopqrstuvwxyz.
\multibb   ABCDEFGHIJKLMNOPQRSTUVWXYZabcdefghijklmnopqrstuvwxyz.
\multiRM   ABCDEFGHIJKLMNOPQRSTUVWXYZabcdefghijklmnopqrstuvwxyz.

\newcommand{\dotleq}{\buildrel \textstyle  .\over {\smash{\lower
      .2ex\hbox{\ensuremath\leqslant}}\vphantom{=}}}
\newcommand{\dotgeq}{\buildrel \textstyle  .\over {\smash{\lower
      .2ex\hbox{\ensuremath\geqslant}}\vphantom{=}}}

\newcommand{\bM}{\begin{bmatrix}}
\newcommand{\eM}{\end{bmatrix}}
\newcommand{\bSM}{\left[\begin{smallmatrix}}
\newcommand{\eSM}{\end{smallmatrix}\right]}
\renewcommand*\env@matrix[1][*\c@MaxMatrixCols c]{%
  \hskip -\arraycolsep
  \let\@ifnextchar\new@ifnextchar
  \array{#1}}

\newqsymbol{`N}{\mathbb{N}}
\newqsymbol{`R}{\mathbb{R}}
\newqsymbol{`P}{\mathbb{P}}
\newqsymbol{`Z}{\mathbb{Z}}

\newqsymbol{`|}{\mid}
\newqsymbol{`8}{\infty}
\newqsymbol{`1}{\left}
\newqsymbol{`2}{\right}
\newqsymbol{`6}{\partial}
\newqsymbol{`0}{\emptyset}
\newqsymbol{`-}{\leftrightarrow}
\newqsymbol{`<}{\langle}
\newqsymbol{`>}{\rangle}

\DeclarePairedDelimiter\abs{\lvert}{\rvert}

\DeclarePairedDelimiter\Set{\{}{\}}
\newcommand{\imod}[1]{\allowbreak\mkern10mu({\operator@font mod}\,\,#1)}

\newcommand{\threecols}[3]{
\hbox to \textwidth{%
      \normalfont\rlap{\parbox[b]{\textwidth}{\raggedright#1\strut}}%
        \hss\parbox[b]{\textwidth}{\centering#2\strut}\hss
        \llap{\parbox[b]{\textwidth}{\raggedleft#3\strut}}%
    }
}

\newcommand{\reason}[2][\relax]{
  \ifthenelse{\equal{#1}{\relax}}{
    \left(\text{#2}\right)
  }{
    \left(\parbox{#1}{\raggedright #2}\right)
  }
}

\let\SavedDoubleVert\relax
\let\protect\relax
{\catcode`\|=\active
  \xdef\extendvert{\protect\expandafter\noexpand\csname extendvert \endcsname}
  \expandafter\gdef\csname extendvert \endcsname#1{\mskip-5mu \left.%
      \ifx\SavedDoubleVert\relax \let\SavedDoubleVert\|\fi
     \:{\let\|\SetDoubleVert
       \mathcode`\|32768\let|\SetVert
     #1}\:\right.\mskip-5mu}
}
\def\SetVert{\@ifnextchar|{\|\@gobble}
    {\egroup\;\mid@vertical\;\bgroup}}
\def\SetDoubleVert{\egroup\;\mid@dblvertical\;\bgroup}

\begingroup
 \edef\@tempa{\meaning\middle}
 \edef\@tempb{\string\middle}
\expandafter \endgroup \ifx\@tempa\@tempb
 \def\mid@vertical{\middle|}
 \def\mid@dblvertical{\middle\SavedDoubleVert}
\else
 \def\mid@vertical{\mskip1mu\vrule\mskip1mu}
 \def\mid@dblvertical{\mskip1mu\vrule\mskip2.5mu\vrule\mskip1mu}
\fi

\makeatother

\usepackage{ctable}
\usepackage{fouridx}
\usepackage{framed}
\usetikzlibrary{positioning,matrix}

\usepackage{paralist}
\usepackage{enumerate}

\usepackage[normalem]{ulem}

\numberwithin{equation}{section}
\makeatletter
\@addtoreset{equation}{section}

\@addtoreset{Theorem}{section}

\@addtoreset{Lemma}{section}

\@addtoreset{Corollary}{section}

\@addtoreset{Example}{section}

\@addtoreset{Remark}{section}

\@addtoreset{Proposition}{section}

\@addtoreset{Definition}{section}

\@addtoreset{Claim}{section}

\@addtoreset{Subclaim}{Theorem}

\makeatother

{\endMakeFramed}

\newenvironment{ybox}{
	\setlength{\FrameSep}{1.5mm}
	\setlength{\FrameRule}{0mm}
  \MakeFramed {\FrameRestore}}%
{\endMakeFramed}

\newenvironment{gbox}{
	\setlength{\FrameSep}{1.5mm}
\setlength{\FrameRule}{0mm}
  \MakeFramed {\FrameRestore}}%
{\endMakeFramed}

{\endMakeFramed}

 {\endMakeFramed}

\usepackage{enumitem}

\usepackage{etoolbox}

\let\theparentequation\theequation
\patchcmd{\theparentequation}{equation}{parentequation}{}{}

\renewenvironment{subequations}[1][]{
	\refstepcounter{equation}%
	\setcounter{parentequation}{\value{equation}}
	\setcounter{equation}{0}
	\let\parentlabel\label
	\ifx\\#1\\\relax\else\label{#1}\fi
	\ignorespaces
}{%
	\setcounter{equation}{\value{parentequation}}
	\ignorespacesafterend
}

\newcommand*{\nextParentEquation}[1][]{
	\refstepcounter{parentequation}
	\setcounter{equation}{0}
	\ifx\\#1\\\relax\else\parentlabel{#1}\fi
}

\PassOptionsToPackage{breaklinks,letterpaper,hyperindex=true,backref=false,bookmarksnumbered,bookmarksopen,linktocpage,colorlinks,linkcolor=BrickRed,citecolor=OliveGreen,urlcolor=Blue,pdfstartview=FitH}{hyperref}
\usepackage{hyperref}


%% file: eg_src.pgf
\def\u{1.5em}
\tikzstyle{dot}=[circle,draw=gray!80,fill=gray!20,thick,inner
sep=2pt,minimum size=1.5*\u]
\tikzstyle{point}=[draw,circle,minimum size=.2em,inner sep=0, outer sep=.2em]
\tikzstyle{+}=[draw,fill=white,circle,minimum size=.8em,inner sep=0pt]
\tikzstyle{arc}=[->]
\tikzstyle{edge}=[-]
\tikzstyle{light}=[gray]
\begin{tikzpicture}[x=.6em,y=.6em,>=latex]
\scriptsize
\path (90:5*\u) node (1) [dot,label={[label distance=0*\u,light]90:$\RZ_1$}] {1};
\path (-30:5*\u) node (2) [dot,label={[label distance=0*\u,light]-30:$\RZ_2$}] {2};
\path (210:5*\u) node (3) [dot,label={[label distance=0*\u,light]-120:$\RZ_3$}] {3};

\draw[edge] (1) to node [right,label={[label distance=0*\u,light]-120:$\RX_{\op{a}}$}] {} (2);
\draw[edge] (1) to node [left,label={[label distance=0*\u,light]-30:$\RX_{\op{c}}$}] {} (3);
\draw[edge] (2) to node [below,label={[label distance=0*\u,light]90:$\RX_{\op{b}}$}] {} (3);

\path[postaction={decorate,decoration={text along path,raise=.3*\u, text align={center},text={{$c(\Set{1,2})=1$}{}}}}] (1) to node [right] {} (2);
\path[postaction={decorate,decoration={text along path,reverse path,raise=.3*\u, text align={center},text={{$c(\Set{1,3})=5$}{}}}}] (1) to node [left] {} (3);
\path[postaction={decorate,decoration={text along path,reverse path,raise=-1*\u, text align={center},text={{$c(\Set{2,3})=1$}{}}}}] (2) to node [above] {} (3);

\end{tikzpicture}

%% file: eg_I.pgf
\def\u{1.5em}

\tikzstyle{dot}=[circle,draw=gray!80,fill=gray!20,thick,inner
sep=2pt,minimum size=1.5*\u]
\tikzstyle{point}=[draw,circle,minimum size=.2em,inner sep=0, outer sep=.2em]
\tikzstyle{+}=[draw,fill=white,circle,minimum size=.8em,inner sep=0pt]
\tikzstyle{arc}=[->]
\tikzstyle{edge}=[-]
\tikzstyle{light}=[gray]
\begin{tikzpicture}[x=.6em,y=.6em,>=latex]
\scriptsize
\path (90:5*\u) node (1) [dot,label={[label distance=0*\u,light]90:{source}}] {1};
\path (-30:5*\u) node (2) [dot] {2};
\path (210:5*\u) node (3) [dot] {3};

\draw[edge] (1) to node [right] {1} (2);
\draw[arc,blue,bend left] (1) edge node [right] {$\Rm_1$} (2);
\draw[edge] (1) to node [left] {5} (3);
\draw[arc,blue,bend left] (1) edge node [right] {$\Rm_1$} (3);
\draw[edge] (2) to node [above] {1} (3);
\draw[arc,purple,bend right] (1) edge node [left] {$\Rm_2$} (3);
\draw[arc,purple,bend right] (3) edge node [below] {$\Rm_2$} (2);

\end{tikzpicture}

%% file: eg_clusters.pgf
\def\u{1.5em}

\tikzstyle{dot}=[circle,draw=gray!80,fill=gray!20,thick,inner
sep=2pt,minimum size=1.5*\u]
\tikzstyle{point}=[draw,circle,minimum size=.2em,inner sep=0, outer sep=.2em]
\tikzstyle{+}=[draw,fill=white,circle,minimum size=.8em,inner sep=0pt]
\tikzstyle{arc}=[->]
\tikzstyle{edge}=[-]
\tikzstyle{light}=[gray]
\begin{tikzpicture}[x=.6em,y=.6em,>=latex]
\scriptsize
\path (90:5*\u) node (1) [dot] {1};
\path (-30:5*\u) node (2) [dot] {2};
\path (210:5*\u) node (3) [dot] {3};

\draw[edge] (1) to node [right] {1} (2);
\draw[edge] (1) to node [left] {5} (3);
\draw[edge] (2) to node [above] {1} (3);
\draw[purple,postaction={decorate,decoration={text along path,text align={left, left indent=4em},reverse path,raise=0.2*\u,text={|\color{purple}|{$`g\in[2,5)$}{}}}}] \convexpath{3,1}{1.5*\u};
\draw[blue,postaction={decorate,decoration={text along path,text align={left, left indent=25em},reverse path,raise=0.2*\u,text={|\color{blue}|{$`g\in[-`8,2)$}{}}}}] \convexpath{1,2,3}{1.8*\u};
\end{tikzpicture}

%% file: eg_orient.pgf
\def\u{1.5em}

\tikzstyle{dot}=[circle,draw=gray!80,fill=gray!20,thick,inner
sep=2pt,minimum size=1.5*\u]
\tikzstyle{point}=[draw,circle,minimum size=.2em,inner sep=0, outer sep=.2em]
\tikzstyle{+}=[draw,fill=white,circle,minimum size=.8em,inner sep=0pt]
\tikzstyle{arc}=[->]
\tikzstyle{edge}=[-]
\tikzstyle{light}=[gray]
\begin{tikzpicture}[x=.6em,y=.6em,>=latex]
\scriptsize
\path (90:5*\u) node (1) [dot] {1};
\path (-30:5*\u) node (2) [dot] {2};
\path (210:5*\u) node (3) [dot] {3};

\draw[arc,postaction={decorate,decoration={text along path,raise=.3*\u, text align={center},text={{$c(1,2)=1$}{}}}}] (1) to node [right] {} (2);
\draw[arc,postaction={decorate,decoration={text along path,reverse path,raise=.3*\u, text align={center},text={{$c(1,3)=5$}{}}}}] (1) to node [left] {} (3);
\draw[arc,postaction={decorate,decoration={text along path,reverse path,raise=-1*\u, text align={center},text={{$c(2,3)=1$}{}}}}] (2) to node [above] {} (3);
\end{tikzpicture}

%% file: eg_cut.pgf
\def\u{1.5em}

\tikzstyle{dot}=[circle,draw=gray!80,fill=gray!20,thick,inner
sep=2pt,minimum size=1.5*\u]
\tikzstyle{point}=[draw,circle,minimum size=.2em,inner sep=0, outer sep=.2em]
\tikzstyle{+}=[draw,fill=white,circle,minimum size=.8em,inner sep=0pt]
\tikzstyle{arc}=[->]
\tikzstyle{edge}=[-]
\tikzstyle{light}=[gray]
\begin{tikzpicture}[x=.6em,y=.6em,>=latex]
\scriptsize
\path (90:5*\u) node (1) [dot] {1};
\path (-30:5*\u) node (2) [dot] {2};
\path (210:5*\u) node (3) [dot] {3};

\draw[arc] (1) to node [right] {1} (2);
\draw[arc] (1) to node [left] {5} (3);
\draw[arc] (2) to node [above] {1} (3);
\node[circle,draw=orange!30!black!70,minimum size=3*\u,label={[orange!30!black!70]below:{$g(\Set{3})=6$}}] at (2) {};
\draw[purple,postaction={decorate,decoration={text along path,text align={left, left indent=10em},raise=0.3*\u,text={|\color{purple}|{$g(\Set{1,3})=1$}{}}}}] \convexpath{3,1}{1.1*\u};
\draw[blue,postaction={decorate,decoration={text along path,text align={left, left indent=30em},raise=0.3*\u,text={|\color{blue}|{$g(\Set{2,3})=6$}{}}}}] \convexpath{2,3}{1.3*\u};
\node[circle,draw=green!30!black!70,minimum size=3*\u,label={[green!30!black!70]below:{$g(\Set{3})=6$}}] at (3) {};
\end{tikzpicture}

%% file: eg_PSP.pgf
\def\u{1.2em}\scriptsize
				\tikzstyle{point}=[draw,circle,minimum size=.2em,inner sep=0, outer sep=0]
\tikzstyle{line}=[gray,thick,dotted]
				\begin{tikzpicture}[x=1.4em,y=1.4em,>=latex]
				\draw[->] (0,-11*\u) -- (0,2*\u) node [label=above:$\hat{g}_{`g}(V)$] {};
				\draw[->] (0,0) -- (6*\u,0) node [label=right:$`g$] {};
				\foreach \i/\ya/\xa/\yb/\xb/\lp/\lb in {
					1/0/0/-6/6/left/{}, 
					2/1/0.5/-10/6/left/{}, 
					3/1/2.5/-6/6/45/{}, 
					5/1/2/-11/6/above/{}} 
				\draw[line] (\xa*\u,\ya*\u)  node [inner sep=0,outer sep=0] {} -- (\xb*\u,\yb*\u);
				\path[gray] (2.5*\u,1*\u)  node [inner sep=0,outer sep=0] {} to node [right,xshift=-0.6*\u,yshift=0.3*\u] {$\begin{aligned}[t] &g_{`g}[\Set{\Set{1,2},\Set{3}}]\\ &=g_{`g}[\Set{\Set{1},\Set{2,3}}] \\ &\kern1em=6-2`g\end{aligned}$} (6*\u,-6*\u);
				\path (2*\u,-2*\u) node (1) [point,red,thick,label=left:{}] {};
				\path (5*\u,-8*\u) node (2) [point,red,thick,label=left:{}] {};
				\node (0) at (0,1*\u) {};
				\draw[-,thick,blue] (0,0) %
				to node [left,yshift=-1*\u] {$g_{`g}[\underbrace{\Set{\Set{1,2,3}}}_{\mcP_0}]=1-`g$} (1) %
				to node [left,yshift=-1*\u] {$g_{`g}[\underbrace{\Set{\Set{1,3},\Set{2}}}_{\mcP_1}]=2-2`g$} (2) %
				to node [left,yshift=-1*\u] {$g_{`g}[\underbrace{\Set{\Set{1},\Set{2},\Set{3}}}_{\mcP_2}]=5-3`g$} (6*\u,-11*\u);
				\draw[purple,dashed] (1)--(1|-0) node [above] {$`g_1=2$};
				\draw[purple,dashed] (2)--(2|-0) node [above] {$`g_2=5$};
				\end{tikzpicture}

%% file: eg_pm2.pgf
\def\u{2em}\scriptsize
\tikzstyle{dot}=[inner sep=0,outer sep=0]
				\tikzstyle{point}=[draw,circle,minimum size=.2em,inner sep=0, outer sep=0]
\tikzstyle{line}=[gray,thick,dotted]
				\begin{tikzpicture}[x=1.4em,y=1.4em,>=latex]
				\draw[->] (0,-2*\u) -- (0,2.5*\u) node (y) [label=45:$\min\limits_{B\subseteq \Set{1,2}:2\in B} g_{`g}(B)-x_{`g}(B`/\Set{2})$] {};
				\draw[->] (-1*\u,0) -- (7.5*\u,0) node (x) [label=right:$`g$] {};

\draw[dot,line] (-1*\u,2*\u)  node [left,xshift=-.2*\u]  {$1-`g$} -- (1*\u,0*\u) -- (3*\u,-2*\u) ;
\draw[dot,line] (-1*\u,0)  node [above left,xshift=-.2*\u] {$0$} -- (7*\u,0*\u) ;

				\path (1*\u,0*\u) node (1) [point,red,thick,label=left:{}] {};
				\node (0) at (0,-0.5*\u) {};
				\draw[-,dot,thick,blue] (-1*\u,0) -- (1*\u,0) node [below left,yshift=-.4*\u] {$B_0=\Set{1,2}$} -- (1) to node[above right, yshift=-.1*\u] {$B_1=\Set{1}$} (3*\u,-2*\u) ;

\tikzstyle{lastP}=[dotted,thick,draw=gray]
\tikzstyle{curP}=[draw=blue]
\tikzstyle{curB}=[draw=blue]
\newcommand{\PIN}{
\node [dot] (1) at (.5*\u,0) {1};
\node [dot] (2) at (1*\u,-1*\u) {2};
\node [dot] (3) at (0,-1*\u) {3};
}
\begin{scope}[yshift=-2.5*\u]
\node at (-1*\u,0*\u) [left] {$B^*(`g)$};
\begin{scope}[xshift=-.5*\u,yshift=0*\u,prefix node name=B0]
\PIN
\end{scope}
\path[curB] \convexpath{B0 1,B0 2}{.3*\u}; 
\begin{scope}[xshift=1.5*\u,yshift=0*\u,prefix node name=B1]
\PIN
\end{scope}
\node at (B1 2) [curB,circle,minimum size=.6*\u] {}; 
\end{scope}

\begin{scope}[yshift=-4.5*\u]
\node at (-1*\u,0*\u) [left] {$\mcP^*(`g)$};
\begin{scope}[xshift=-.5*\u,yshift=0*\u,prefix node name=P0]
\PIN
\end{scope}
\node at (P0 1) [lastP,circle,minimum size=.8*\u] {}; 
\path[curP] \convexpath{P0 1,P0 2}{.3*\u}; 
\begin{scope}[xshift=1.5*\u,yshift=0*\u,prefix node name=P1]
\PIN
\end{scope}
\node at (P1 1) [lastP,circle,minimum size=.8*\u] {}; 
\node at (P1 1) [curP,circle,minimum size=.6*\u] {}; 
\node at (P1 2) [curP,circle,minimum size=.6*\u] {}; 
\end{scope}

				\path[purple,dashed] (1|-y) + (0,-1*\u) node [above] {$`g'_1=1$} [draw] -- (P1 3.south-|1);

				\end{tikzpicture}

%% file: eg_pm3.pgf
\def\u{2em}\scriptsize
\tikzstyle{dot}=[inner sep=0,outer sep=0]
				\tikzstyle{point}=[draw,circle,minimum size=.2em,inner sep=0, outer sep=0]
\tikzstyle{line}=[gray]
				\begin{tikzpicture}[x=1.4em,y=1.4em,>=latex]
				\draw[->] (0,-1*\u) -- (0,7.5*\u) node (y) [label=45:$\min\limits_{B\subseteq \Set{1,2,3}:3\in B} g_{`g}(B)-x_{`g}(B`/\Set{3})$] {};
				\draw[->] (0,0) -- (7.5*\u,0) node (x) [label=right:$`g$] {};

\draw[dot,line] (0,0)  node [left,xshift=-.2*\u]  {$\max\Set{0,`g-1}$} -- (1*\u,0*\u) -- (7*\u,6*\u) ;
\draw[dot,line] (0,2*\u)  node [left,xshift=-.2*\u] {$\max\Set{2-`g,1}$} -- (1*\u,1*\u) -- (7*\u,1*\u) ;
\draw[dot,line] (0,5*\u) node [left,xshift=-.2*\u] {$5$} to node [above] {$B=\Set{2,3}$} (7*\u,5*\u);
\draw[dot,line] (0,6*\u) node [left,xshift=-.2*\u] {$6-`g$} -- (7*\u,-1*\u) ;

				\path (2*\u,1*\u) node (2) [point,red,thick,label=left:{}] {};
				\path (5*\u,1*\u) node (3) [point,red,thick,label=left:{}] {};
				\node [dot] (0) at (0,0) {};
				\draw[-,dot,thick,blue] (0,0) -- (1*\u,0)  -- (2) to node[below, yshift=-.1*\u] {$B_1=\Set{1,3}$} (3) to node[above right,yshift=.2*\u] {$B_2=\Set{3}$} (7*\u,-1*\u);
				\node [blue,below left,xshift=0*\u,yshift=-0.2*\u] at (2|-0) {$B_0=\Set{1,2,3}$};

\tikzstyle{lastP}=[dotted,thick,draw=gray]
\tikzstyle{curP}=[draw=blue]
\tikzstyle{curB}=[draw=blue]
\newcommand{\PIN}{
\node [dot] (1) at (.5*\u,0) {1};
\node [dot] (2) at (1*\u,-1*\u) {2};
\node [dot] (3) at (0,-1*\u) {3};
}
\begin{scope}[yshift=-2*\u]
\node at (-1*\u,0*\u) [left] {$B^*(`g)$};
\begin{scope}[xshift=0.5*\u,yshift=0*\u,prefix node name=B0]
\PIN
\end{scope}
\path[curB] \convexpath{B0 1,B0 2,B0 3}{.3*\u}; 
\begin{scope}[xshift=2.5*\u,yshift=0*\u,prefix node name=B1]
\PIN
\end{scope}
\path[curB] \convexpath{B1 1,B1 3}{.3*\u}; 
\begin{scope}[xshift=5.5*\u,yshift=0*\u,prefix node name=B2]
\PIN
\end{scope}
\node at (B2 3) [curB,circle,minimum size=.6*\u] {}; 

\end{scope}

\begin{scope}[yshift=-4*\u]
\node at (-1*\u,0*\u) [left] {$\mcP^*(`g)$};
\begin{scope}[xshift=.5*\u,yshift=0*\u,prefix node name=P0]
\PIN
\end{scope}
\path[lastP] \convexpath{P0 1,P0 2}{.4*\u}; 
\node at (P0 1) [lastP,circle,minimum size=.9*\u] {}; 
\node at (P0 2) [lastP,circle,minimum size=.9*\u] {}; 
\path[curP] \convexpath{P0 1,P0 2,P0 3}{.3*\u}; 
\begin{scope}[xshift=2.5*\u,yshift=0*\u,prefix node name=P1]
\PIN
\end{scope}
\node at (P1 1) [lastP,circle,minimum size=.8*\u] {}; 
\node at (P1 2) [lastP,circle,minimum size=.8*\u] {}; 
\path[curP] \convexpath{P1 1,P1 3}{.3*\u}; 
\node at (P1 2) [curP,circle,minimum size=.6*\u] {}; 
\begin{scope}[xshift=5.5*\u,yshift=0*\u,prefix node name=P2]
\PIN
\end{scope}
\node at (P2 1) [lastP,circle,minimum size=.8*\u] {}; 
\node at (P2 2) [lastP,circle,minimum size=.8*\u] {}; 
\node at (P2 1) [curP,circle,minimum size=.6*\u] {}; 
\node at (P2 2) [curP,circle,minimum size=.6*\u] {}; 
\node at (P2 3) [curP,circle,minimum size=.6*\u] {}; 
\end{scope}

\begin{scope}[yshift=-6*\u,blue]
\node (PSP1) at (1*\u,0) {$\mcP_1$};
\node at (3*\u,0) {$\mcP_2$};
\node at (6*\u,0) {$\mcP_3$};

\end{scope}

				\path[purple,dashed] (2|-y) + (0,-1*\u) node [above] {$`g'_1=2$} [draw] -- (PSP1.south-|2) node [below] {$`g_1$};
				\path[purple,dashed] (3|-y) + (0,-1*\u) node [above] {$`g'_2=5$} [draw] -- (PSP1.south-|3) node [below] {$`g_2$};
				\end{tikzpicture}

%% file: eg_D2.pgf
\def\u{1em}

\tikzstyle{dot}=[circle,draw=gray!80,fill=gray!20,thick,inner
sep=2pt,minimum size=1.5*\u]
\tikzstyle{point}=[draw,circle,minimum size=.2em,inner sep=0, outer sep=.2em]
\tikzstyle{+}=[draw,fill=white,circle,minimum size=.8em,inner sep=0pt]
\tikzstyle{arc}=[->]
\tikzstyle{edge}=[-]
\tikzstyle{light}=[gray]
\begin{tikzpicture}[x=.6em,y=.6em,>=latex]
\scriptsize
\path (90:5*\u) node (1) [dot] {1};
\path (-30:5*\u) node (2) [dot] {2};
\path (1) +(-9*\u,0*\u) node [dot] (s) {s};

\draw[arc] (1) to node [right,xshift=-.8*\u] {$\begin{aligned}
&\abs{x_{`g,1}}^++1\\[-.2em]  &\kern1em=\begin{cases}1-`g, & `g<0\\ 1, & `g\geq 0\end{cases}
\end{aligned}$} (2);

\draw[arc] (s) to node [above,xshift=-1*\u] {$\begin{aligned}
&\kern-.8em\abs{-x_{`g,1}}^+\\[-.7em]  &\kern1em=\begin{cases}
0, & `g<0\\ 
`g, & `g\geq 0\end{cases}
\end{aligned}$} (1);


\end{tikzpicture}

%% file: eg_D2_contract.pgf
\def\u{1em}

\tikzstyle{dot}=[circle,draw=gray!80,fill=gray!20,thick,inner
sep=2pt,minimum size=1.5*\u]
\tikzstyle{point}=[draw,circle,minimum size=.2em,inner sep=0, outer sep=.2em]
\tikzstyle{+}=[draw,fill=white,circle,minimum size=.8em,inner sep=0pt]
\tikzstyle{arc}=[->]
\tikzstyle{edge}=[-]
\tikzstyle{light}=[gray]
\begin{tikzpicture}[x=.6em,y=.6em,>=latex]
\scriptsize
\path (90:5*\u) node (1) [dot] {1};
\path (-30:5*\u) node (2) [dot] {2};
\path (210:5*\u) node (3) [dot,dotted] {3};
\path (1) +(-9*\u,0*\u) node [dot] (s) {s};
\path (1) +(9*\u,0*\u) node [dot,dotted] (t) {t};

\draw[arc] (1) to node [right] {$1$} (2);
\draw[arc,red,dotted] (1) to node [left] {$5$} (3);
\draw[arc,red,dotted] (2) to node  [above] {$1$} (3);

\draw[blue,arc] (s) to node [above,yshift=.4*\u] {$\begin{aligned}
&\abs{x_{`g,1}}^+\\[-.7em]  &\kern1em=\begin{cases}0, & `g<0\\ `g, & `g\geq 0\end{cases}
\end{aligned}$} (1);
\draw[arc,blue] (1) to node [above,yshift=.4*\u] {$
\begin{aligned}
&\kern-.8em\abs{-x_{`g,1}}^+\\[-.7em]  &\kern1em=\begin{cases}
-`g, & `g<0\\ 
0, & `g\geq 0\end{cases}
\end{aligned}$} (t);

\draw[blue] \convexpath{s,3}{1*\u};
\draw[blue] \convexpath{2,t}{1*\u};

\end{tikzpicture}

%% file: eg_D2_1.pgf
\def\u{1em}

\tikzstyle{dot}=[circle,draw=gray!80,fill=gray!20,thick,inner
sep=2pt,minimum size=1.5*\u]
\tikzstyle{point}=[draw,circle,minimum size=.2em,inner sep=0, outer sep=.2em]
\tikzstyle{+}=[draw,fill=white,circle,minimum size=.8em,inner sep=0pt]
\tikzstyle{arc}=[->]
\tikzstyle{edge}=[-]
\tikzstyle{light}=[gray]
\begin{tikzpicture}[x=.6em,y=.6em,>=latex]
\scriptsize
\path (90:5*\u) node (1) [dot] {1};
\path (-30:5*\u) node (2) [dot] {2};
\path (1) +(-9*\u,0*\u) node [dot] (s) {s};

\draw[arc] (1) to node [right] {$1$} (2);

\draw[arc] (s) to node [above] {$1$} (1);

\draw[purple] \convexpath{1,2}{1*\u} node [left,yshift=-2*\u,xshift=1*\u] {$c(s,\Set{1,2})=1$};
\node[minimum size=2.5*\u,draw=purple,circle] at (2) [label={[purple]left:{$c(\Set{s,1},2)=1$}}] {}; 

\end{tikzpicture}

%% file: eg_D3.pgf
\def\u{1em}

\tikzstyle{dot}=[circle,draw=gray!80,fill=gray!20,thick,inner
sep=2pt,minimum size=1.5*\u]
\tikzstyle{point}=[draw,circle,minimum size=.2em,inner sep=0, outer sep=.2em]
\tikzstyle{+}=[draw,fill=white,circle,minimum size=.8em,inner sep=0pt]
\tikzstyle{arc}=[->]
\tikzstyle{edge}=[-]
\tikzstyle{light}=[gray]
\begin{tikzpicture}[x=.6em,y=.6em,>=latex]
\scriptsize
\path (90:5*\u) node (1) [dot] {1};
\path (-30:5*\u) node (2) [dot] {2};
\path (210:5*\u) node (3) [dot] {3};
\path (1) +(9*\u,0*\u) node [dot] (s) {s};

\draw[arc] (1) to node [right] {$1$} (2);
\draw[arc] (1) to node [left] {$5$} (3);
\draw[arc] (2) to node  [xshift=2*\u,below] {$
\begin{aligned}
&\kern-.8em\abs{x_{`g,2}}^++1\\[-.7em]  &\kern1em=\begin{cases}
2-`g, & `g<1\\ 
1, & `g\geq 1\end{cases}
\end{aligned}$} (3);

\draw[arc] (s) to node [above,yshift=.4*\u] {$\begin{aligned}
&\abs{-x_{`g,1}}^+\\[-.7em]  &\kern1em=\begin{cases}1, & `g<1\\ `g, & `g\geq 1\end{cases}
\end{aligned}$} (1);
\draw[arc] (s) to node [pos=.6,right,xshift=.8*\u] {$\begin{aligned}
&\abs{-x_{`g,2}}^+\\[0em]  &\kern-1.8em=\begin{cases}0, & `g<1\\ `g-1, & `g\geq 1\end{cases}\kern-2em
\end{aligned}$} (2);


\end{tikzpicture}

%% file: eg_D3_contract.pgf
\def\u{1em}

\tikzstyle{dot}=[circle,draw=gray!80,fill=gray!20,thick,inner
sep=2pt,minimum size=1.5*\u]
\tikzstyle{point}=[draw,circle,minimum size=.2em,inner sep=0, outer sep=.2em]
\tikzstyle{+}=[draw,fill=white,circle,minimum size=.8em,inner sep=0pt]
\tikzstyle{arc}=[->]
\tikzstyle{edge}=[-]
\tikzstyle{light}=[gray]
\begin{tikzpicture}[x=.6em,y=.6em,>=latex]
\scriptsize
\path (90:5*\u) node (1) [dot] {1};
\path (-30:5*\u) node (2) [dot] {2};
\path (210:5*\u) node (3) [dot] {3};
\path (3) +(-3*\u,0*\u) node [dot,dotted] (t) {t};
\path (1) +(9*\u,0*\u) node [dot] (s) {s};

\draw[arc] (1) to node [right] {$1$} (2);
\draw[arc] (1) to node [left] {$5$} (3);
\draw[arc] (2) to node  [above] {$1$} (3);

\draw[arc,blue] (s) to node [above,yshift=.4*\u] {$\begin{aligned}
&\abs{-x_{`g,1}}^+\\[-.7em]  &\kern1em=\begin{cases}1, & `g<1\\ `g, & `g\geq 1\end{cases}
\end{aligned}$} (1);
\draw[arc,blue] (s) to node [pos=.6,right,xshift=.8*\u] {$\begin{aligned}
&\abs{-x_{`g,2}}^+\\[0em]  &\kern-1.8em=\begin{cases}0, & `g<1\\ `g-1, & `g\geq 1\end{cases}\kern-3em
\end{aligned}$} (2);
\draw[arc,blue,bend left] (2) edge node [below,yshift=.4*\u] {$
\begin{aligned}
&\kern-.8em\abs{x_{`g,2}}^+\\[-.7em]  &\kern1em=\begin{cases}
1-`g, & `g<1\\ 
0, & `g\geq 1\end{cases}
\end{aligned}$} (t);

\draw[blue] \convexpath{t,3}{1*\u};
\node[draw=blue,circle,minimum size=2*\u] at (s) {};

\end{tikzpicture}

%% file: eg_D3_1.pgf
\def\u{1em}

\tikzstyle{dot}=[circle,draw=gray!80,fill=gray!20,thick,inner
sep=2pt,minimum size=1.5*\u]
\tikzstyle{point}=[draw,circle,minimum size=.2em,inner sep=0, outer sep=.2em]
\tikzstyle{+}=[draw,fill=white,circle,minimum size=.8em,inner sep=0pt]
\tikzstyle{arc}=[->]
\tikzstyle{edge}=[-]
\tikzstyle{flow}=[green!30!black!70!]
\tikzstyle{light}=[gray]
\begin{tikzpicture}[x=.6em,y=.6em,>=latex]
\scriptsize

\begin{scope}
\path (90:5.8*\u) node (1) [dot] {1};
\path (-30:5.8*\u) node (2) [dot] {2};
\path (210:5.8*\u) node (3) [dot] {3};
\path (1) +(9*\u,0*\u) node [dot] (s) {s};

\draw[arc,postaction={decorate,decoration={text along path,raise=-.8*\u, text align={center},text={|\color{green!30!black!70}|{$(f^*(1,2)=0)$}{}}}}] (1) to node [pos=.6,right] {$1$} node [pos=.6,left,flow] {} (2);
\draw[arc,postaction={decorate,decoration={text along path,raise=-.8*\u,reverse path, text align={center},text={|\color{green!30!black!70}|{$(f^*(1,3)=1)$}{}}}}] (1) to node [left] {$5$} node [right,flow] {} (3);
\draw[arc,postaction={decorate,decoration={text along path,raise=0.3*\u,reverse path, text align={center},text={|\color{green!30!black!70}|{$(f^*(2,3)=0)$}{}}}}] (2) to node  [below] {$1$} node  [above,flow] {} (3);

\draw[arc] (s) to node [above] {$1$} node [flow,below] {$(f^*(s,1)=1)$} (1);

\draw[purple] \convexpath{1,2,3}{1*\u} node [left] {$T^*=\Set{1,2,3}$};
\end{scope}

\end{tikzpicture}

%% file: eg_D3_35.pgf
\def\u{1em}

\tikzstyle{dot}=[circle,draw=gray!80,fill=gray!20,thick,inner
sep=2pt,minimum size=1.5*\u]
\tikzstyle{point}=[draw,circle,minimum size=.2em,inner sep=0, outer sep=.2em]
\tikzstyle{+}=[draw,fill=white,circle,minimum size=.8em,inner sep=0pt]
\tikzstyle{arc}=[->]
\tikzstyle{edge}=[-]
\tikzstyle{flow}=[green!30!black!70!]
\tikzstyle{light}=[gray]
\begin{tikzpicture}[x=.6em,y=.6em,>=latex]
\scriptsize

\begin{scope}
\path (90:5*\u) node (1) [dot] {1};
\path (-30:5*\u) node (2) [dot] {2};
\path (210:5*\u) node (3) [dot] {3};
\path (1) +(9*\u,0*\u) node [dot] (s) {s};

\draw[arc] (1) to node [pos=.6,right] {$1$} node [pos=.6,left,flow] {$(0)$} (2);
\draw[arc] (1) to node [left] {$5$} node [pos=.4,right,flow] {$(1)$} (3);
\draw[arc] (2) to node  [below] {$1$} node  [above,flow] {$(0)$} (3);

\draw[arc] (s) to node [above] {$3.5$} node [flow,below] {$(1)$} (1);
\draw[arc] (s) to node [right] {$2.5$} node [flow,left] {$(0)$} (2);

\node[draw=blue,circle,minimum size=3*\u,label={[blue]right:$S$}] at (s) {};
\node[draw=blue,circle,minimum size=3*\u,label={[blue]left:$T$}] at (3) {};

\end{scope}

\begin{scope}[yshift=-12*\u]
\path (90:5*\u) node (1) [dot] {1};
\path (-30:5*\u) node (2) [dot] {2};
\path (210:5*\u) node (3) [dot] {3};
\path (1) +(9*\u,0*\u) node [dot] (s) {s};

\draw[arc] (1) to node [pos=.6,right] {$1$} node [pos=.7,left,flow] {$(0)$} (2);
\draw[arc] (1) to node [left] {$5$} node [pos=.45,right,flow] {$(3.5)$} (3);
\draw[arc] (2) to node  [below] {$1$} node  [above,flow] {$(1)$} (3);

\draw[arc] (s) to node [above] {$3.5$} node [flow,below] {$(3.5)$} (1);
\draw[arc] (s) to node [right] {$2.5$} node [flow,left] {$(1)$} (2);


\draw[purple] \convexpath{1,3}{1*\u} node [left] {$T^*=\Set{1,3}$};
\end{scope}

\end{tikzpicture}

%% file: eg_D3_2.pgf
\def\u{1em}

\tikzstyle{dot}=[circle,draw=gray!80,fill=gray!20,thick,inner
sep=2pt,minimum size=1.5*\u]
\tikzstyle{point}=[draw,circle,minimum size=.2em,inner sep=0, outer sep=.2em]
\tikzstyle{+}=[draw,fill=white,circle,minimum size=.8em,inner sep=0pt]
\tikzstyle{arc}=[->]
\tikzstyle{edge}=[-]
\tikzstyle{flow}=[green!30!black!70!]
\tikzstyle{light}=[gray]
\begin{tikzpicture}[x=.6em,y=.6em,>=latex]
\scriptsize

\begin{scope}
\path (90:5*\u) node (1) [dot] {1};
\path (-30:5*\u) node (2) [dot] {2};
\path (210:5*\u) node (3) [dot] {3};
\path (1) +(9*\u,0*\u) node [dot] (s) {s};

\draw[arc] (1) to node [pos=.6,right] {$1$} node [pos=.6,left,flow] {$(0)$} (2);
\draw[arc] (1) to node [left] {$5$} node [right,flow] {$(1)$} (3);
\draw[arc] (2) to node  [below] {$1$} node  [above,flow] {$(0)$} (3);

\draw[arc] (s) to node [above] {$2$} node [flow,below] {$(1)$} (1);
\draw[arc] (s) to node [right] {$1$} node [flow,left] {$(0)$} (2);

\node[draw=blue,circle,minimum size=3*\u,label={[blue]right:$S$}] at (s) {};
\path[draw=blue] \convexpath{1,3}{1.5*\u} node [blue,left] {$T$};

\end{scope}

\begin{scope}[yshift=-12*\u]
\path (90:5*\u) node (1) {};
\path (-30:5*\u) node (2) [dot] {2};
\path (210:5*\u) node (3) [dot] {3};
\path (1) +(9*\u,0*\u) node [dot] (s) {s};

\draw[arc] (2) to node  [below] {$1$} node  [above,flow] {$(1)$} (3);

\draw[arc] (s) to node [above] {$2$} node [flow,below] {$(2)$} (3);
\draw[arc] (s) to node [right] {$1$} node [flow,left] {$(1)$} (2);

\node[draw=purple,circle,minimum size=2*\u,label={[purple]above:$T^*=\Set{3}$}] at (3) {};

\end{scope}

\end{tikzpicture}

%% file: eg_D3_5.pgf
\def\u{1em}

\tikzstyle{dot}=[circle,draw=gray!80,fill=gray!20,thick,inner
sep=2pt,minimum size=1.5*\u]
\tikzstyle{point}=[draw,circle,minimum size=.2em,inner sep=0, outer sep=.2em]
\tikzstyle{+}=[draw,fill=white,circle,minimum size=.8em,inner sep=0pt]
\tikzstyle{arc}=[->]
\tikzstyle{edge}=[-]
\tikzstyle{flow}=[green!30!black!70!]
\tikzstyle{light}=[gray]
\begin{tikzpicture}[x=.6em,y=.6em,>=latex]
\scriptsize

\begin{scope}
\path (90:5*\u) node (1) [dot] {1};
\path (-30:5*\u) node (2) [dot] {2};
\path (210:5*\u) node (3) [dot] {3};
\path (1) +(9*\u,0*\u) node [dot] (s) {s};

\draw[arc] (1) to node [pos=.6,right] {$1$} node [pos=.7,left,flow] {$(0)$} (2);
\draw[arc] (1) to node [left] {$5$} node [pos=.45,right,flow] {$(3.5)$} (3);
\draw[arc] (2) to node  [below] {$1$} node  [above,flow] {$(1)$} (3);

\draw[arc] (s) to node [above] {$5$} node [flow,below] {$(3.5)$} (1);
\draw[arc] (s) to node [right] {$4$} node [flow,left] {$(1)$} (2);

\node[draw=blue,circle,minimum size=3*\u,label={[blue]left:$T$}] at (3) {};
\path[draw=blue] \convexpath{s,2}{1.5*\u} node [blue,left] {$S$};
\end{scope}

\begin{scope}[yshift=-12*\u]
\path (90:5*\u) node (1) [dot] {1};
\path (210:5*\u) node (3) [dot] {3};
\path (1) +(9*\u,0*\u) node [dot] (s) {s};

\draw[arc] (1) to node [left] {$5$} node [right,flow] {$(5)$} (3);
\draw[arc] (s) to node  [below] {$1$} node  [above,flow] {$(1)$} (3);

\draw[arc] (s) to node [above] {$5$} node [flow,below] {$(5)$} (1);

\node[draw=purple,circle,minimum size=2*\u,label={[purple]right:$T^*=\Set{3}$}] at (3) {};
\end{scope}

\end{tikzpicture}